\documentclass[11pt,preprintnumbers,pra,aps,
nofootinbib,tightenlines,floatfix, twocolumn,superscriptaddress,longbibliography]{revtex4-2}

\usepackage[colorlinks=true]{hyperref}
\hypersetup{
  linkcolor  = TealBlue!90!Black,
  citecolor  = YellowGreen!90!Black,
  urlcolor   = Purple!90!Black,
  colorlinks = true,
}
\usepackage[dvipsnames]{xcolor}
\usepackage[all]{hypcap}
\usepackage{bbm}
\usepackage{amsmath,amssymb}
\usepackage{graphicx}
\usepackage{physics}
\usepackage{balance}
\usepackage[utf8]{inputenc}
\usepackage{csquotes}
\usepackage{empheq}
\usepackage{longtable}
\usepackage{multirow}
\usepackage[section]{placeins}
\usepackage{comment}
\usepackage[customcolors]{hf-tikz}
\usepackage[top=1in,bottom=1in,left=1.9cm,right=1.9cm]{geometry}
\usepackage{setspace}
\usepackage{mathtools}
\usepackage{qcircuit}
\usepackage{cancel}

\usepackage{tikz}
\usetikzlibrary{decorations.pathreplacing,calc}
\newcommand{\tikzmark}[1]{\tikz[overlay,remember picture] \node (#1) {};}

\allowdisplaybreaks
\newcommand\scalemath[2]{\scalebox{#1}{\mbox{\ensuremath{\displaystyle #2}}}}

\catcode`,\active

\catcode`\,12

\begin{document}

\title{Deforming the Trail: Baseline Quantum Circuitry for \texorpdfstring{$\text{SU(2)}_k$}{SU(2)k} Lattice Gauge Theory}

\author{Zoë Webb-Mack}
\email{zoe.webb-mack@duke.edu}
\affiliation{{Duke Quantum Center and Department of Physics, Duke University, Durham, NC 27708, USA}}
\author{Natalie Klco}
\email{natalie.klco@duke.edu}
\affiliation{{Duke Quantum Center and Department of Physics, Duke University, Durham, NC 27708, USA}}

\begin{abstract}
Quantifying quantum resources for simulating the fundamental forces of Nature is sensitive to the mapping of gauge fields onto finite quantum computational architectures.
When locally truncating lattice gauge theories in the irreducible representation basis, it has been proposed to further deform the theory via quantum groups.
The purpose of this deformation is (1) to provide an infinite tower of finite-dimensional ($d = k+1$) groups systematically approximating the infinite-dimensional gauge links and (2) to restore the physical unitarity of a plaquette operator diagonalization procedure analytically derived from the field continuum by recontracting vertex pairs.  
For the SU(2)$_k$ Yang-Mills pure-gauge theory, we provide a constructive strategy of gauge-variant completions to extend this unitarity to the entire computational Hilbert space, leading to well-defined time evolution unitaries as targets for optimized circuit synthesis.
Leveraging basic circuit decompositions and symmetries of the diagonalized plaquette operator, we report resource upper-bounds on the generalized-controlled-X two-qudit gates for arbitrary local truncation~$d$, reducing estimates and scaling relative to the non-deformed theory by three polynomial powers from $\mathcal{O}(d^8)$ to $\mathcal{O}(d^5)$.
Examining the stronger q-deformed gauge constraint, which softens the total flux at vertices, we show that the physical Hilbert space dimension of the deformed plaquette operator scales equivalently to its non-deformed counterpart with a constant factor $0.2563(5)$.
Thus, despite affecting interactions at all scales as exemplified by the observed flux hierarchy inversion symmetry, q-deformation continues to pass scrutiny as a reliable truncation offering advantages in quantum circuit synthesis.
\end{abstract}

\date{\today}
\maketitle

{
\footnotesize
\tableofcontents
}

\section{Introduction}
Quantum simulation of lattice gauge theories (LGTs) has emerged as a promising avenue for accessing non-perturbative and dynamical regimes of the Standard Model. Beyond discretization of space, further truncating the infinite-dimensional bosonic Hilbert spaces of the gauge field links allows their representation on finite quantum computational devices~\cite{Davoudi:2020yln,Banuls:2019bmf, Aidelsburger:2021mia, Klco:2021lap, Bauer:2022hpo, Bauer:2023qgm}.  Currently pursued representations of LGTs for efficient implementation on quantum devices are vast, including group element basis approaches~\cite{Zohar:2014qma}, loop-string-hadron formulation~\cite{Raychowdhury:2019iki}, discretized light-front formulation~\cite{Kreshchuk:2020dla}, quantum link models~\cite{Wiese:2021djl}, finite subgroup approaches \cite{Alexandru:2021jpm, Ji:2020kjk, Gonzalez-Cuadra:2022hxt}, and representation basis cutoffs \cite{Byrnes:2005qx,
Ciavarella:2021nmj, Klco:2019evd}. 
As these and other approaches are explored in increasing detail~\cite{Zohar:2012xf, Banerjee:2012xg, Zohar:2013zla, Zohar:2015hwa, Zohar:2018cwb, Kasper:2020akk, Buser:2020cvn, Bauer:2021gup,Bauer:2021gek, Kan:2021xfc, Haase:2020kaj, Kadam:2022ipf, Davoudi:2022xmb, Pardo:2022hrp, Liu:2023lsr, Ciavarella:2023mfc, DAndrea:2023qnr, Gustafson:2024kym, Rhodes:2024zbr, Ciavarella:2024fzw, Grabowska:2024emw, Kavaki:2024ijd, Halimeh:2024bth, Ciavarella:2025bsg, Balaji:2025afl, Perez:2025cxl}, the set as a whole provides simulation design flexibility mirroring the ongoing evolution in quantum architectures.

In an electric basis with higher than one spatial dimension, implementing the time evolution operator associated with the magnetic term of the Hamiltonian is a distinct challenge due to its multi-body interaction of four (or more) gauge links.
In this case, truncating with finite-dimensional gauge groups, such as those employed in finite subgroup approaches, can lead to strategic circuit synthesis as they admit transforms with efficient circuit implementations~\cite{Bacon:2004moa,Murairi:2024xpc}. 
However, a finite subgroup approach to truncation can introduce additional phases into the LGT phase diagram, challenging their ability to accurately capture the field continuum~\cite{Bhanot:1981xp}. 
This is of particular concern for non-Abelian Lie groups, where the presence of a maximal finite subgroup can inhibit the ability to raise the truncation and suppress digitization artifacts.
For SU(2), a maximal finite subgroup may be sufficient to support continuum simulations~\cite{PhysRevD.22.2465, Alexandru:2019nsa}. 
However for SU(3), none of the available finite subgroups are expected to be compatible with the continuum~\cite{Bhanot:1981xp} without modified interactions~\cite{BHANOT1982337,Perez:2025cxl, Ji:2020kjk, Alexandru:2019nsa, Alexandru:2021jpm, Assi:2024pdn}. 

Rather than extend the action to improve convergence with a finite subgroup of the theory, q-deformation modifies the theory itself to provide a closed symmetry group at every truncation~\cite{Kirillov:1991ec, Biedenharn1995, Kassel, 1998RvMaP..10..511P,Levin:2004mi,Bonesteel:2012pkv,Hayata:2023puo}.
In doing so, the limited convergence of maximal finite subgroups is addressed by providing an infinite tower of finite groups that converge to the non-deformed gauge theory at high local dimension. 
Here, we focus on q-deformed quantum simulation for SU(2) pure gauge theory~\cite{Zache:2023dko,Hayata:2026rmv,John:2026gut} as a benchmark whose techniques can be extended to SU(3).

The full plaquette operator can be diagonalized using a sequence of \textit{F-moves}~\cite{Robson1982}, which alter the contractions of angular momenta at neighboring vertices and shrink the active space of the plaquette operator to a single link. 
By enabling an infinite tower of finite groups, q-deformation preserves the unitarity of F-moves in the truncated gauge-invariant (physical) subspace, allowing this sequence to be available in truncated theories.
For pure state quantum simulation, additional gauge-variant completion (GVC)~\cite{Klco:2019evd} is required to complete the F-move unitarily over the remaining gauge-variant (unphysical) subspace of the computational Hilbert space. 
In the present work, we complete the diagonalizing sequence of F-moves with GVCs in two stages: once during the diagonalization in order to reduce the size of the plaquette operator, enabling a resource advantage over circuit synthesis strategies without diagonalization, and once when identifying gates for device implementation.

In Section~\ref{sec: Main: YM theory} we describe the SU(2) Yang-Mills theory and write the Hamiltonian in terms of \textit{F-symbols}. In Section~\ref{sec: Main: non-deformed F-moves}, we sketch the sequence of F-moves that diagonalizes a single plaquette operator in a plaquette chain. In Section~\ref{sec: q-deformed F-moves}, supported by Appendix~\ref{sec: Appendix: F def and props}, we state the q-deformed analog of the F-move and discuss the q-deformed gauge-invariant subspace, which is further constrained to limit the total flux at vertices. 
Using transfer matrix techniques in Section~\ref{sec: Main: convergence}, our analysis shows that the q-deformed single-plaquette gauge-invariant subspace still grows towards the continuous field at the same rate as the non-deformed theory.
This observation is consistent with continuum convergence properties being retained upon q-deformation.

In Section~\ref{sec: qdefDiagF1}, supported by Appendix~\ref{sec: Appendix: diagonalization}, we apply our first GVC to derive a sequence of q-deformed phased F-moves that diagonalize the SU(2)$_k$ Yang-Mills plaquette operator on a 1D plaquette chain and shrink its active space to a single link. This provides the core content applicable to higher spatial dimensions. Following this F-sequence, the transformed plaquette operator is observed to have an additional symmetry under inversion of low and high flux in Section~\ref{sec: Main: FHI}.
In Section~\ref{sec:circuitStrategy}, we present a scalable and diagrammatic strategy for synthesizing unitaries with a suitable choice of a second GVC, connecting circuit elements to angular momentum contractions satisfying Gauss's law. 
As a demonstration of this strategy, circuit content for the two lowest truncations (qubit and qutrit) are provided in Appendix~\ref{sec: Appendix: circuit implementation}. 
In Section~\ref{sec: Main: resource scaling}, we upper-bound two-qudit entangling gate resource estimates of the diagonalized plaquette time evolution operator, achieving a reduction in resource scaling compared to non-deformed techniques~\cite{Jiang:2025ufg}. 
By providing a complete and concrete simulation protocol for q-deformed pure-gauge theories, this work supports future  optimizations (such as those described in Appendices~\ref{Appendix: counting actively mixed},~\ref{Appendix: circuits}, and \ref{sec: Appendix: decomp}) in collaboration with quantum hardware and software developments.

\section{SU(2) Yang Mills theory on the lattice}\label{sec: Main: YM theory}

Consider a pure-gauge SU(2) Yang-Mills theory on a one-dimensional plaquette chain with periodic boundary conditions (PBCs). With unit lattice spacing and coupling $g^2$, the Hamiltonian for the theory is~\cite{Kogut:1974ag}
\begin{equation}\label{eq: Hamiltonian}
    H = \frac{g^2}{2}\sum_{l}E_l^2 - \frac{1}{2g^2}\sum_{\text{plaq's}}(\Box + \Box^\dagger) \ \ \ ,
\end{equation}
where $E_l^2$ is the Casimir invariant of the group (summed over all links $l$ constituting the electric term of the Hamiltonian) and $\Box$ is the plaquette operator (summed over all plaquettes constituting the magnetic term). 
We work in the electric representation basis, associating irrep and left- and right-projection degrees of freedom ($|j_lm_lm_l'\rangle$, respectively) to each link $l$. 
For fixed $k \in \mathbb{N}$, we impose angular momentum truncation $k/2$ as the maximal $j$ value for each link, i.e., each local Hilbert space spans $j \in \{0, 1/2, ... k/2\}$. In this electric basis, the electric operator $E^2$ is diagonal~\cite{Zohar:2016iic, Robson1982}, while the plaquette operator has a non-trivial structure due to its non-commutivity with $E^2$. 
In this work, we are interested in identifying a unitary transformation from the electric basis to a basis of plaquette operator eigenstates. 
Because the eigenbasis of $e^{-i\tau \Box}$ is the same as that of $\Box$, this transformation likewise diagonalizes the plaquette operator time evolution.  This renders a quantum simulation strategy that alternates between canonically conjugate eigenbases, e.g., familiar from scalar fields~\cite{Jordan:2011ci,Jordan:2012xnu,Klco:2018zqz}.

\begin{figure*}
    \centering
    \includegraphics[width=0.9\linewidth]{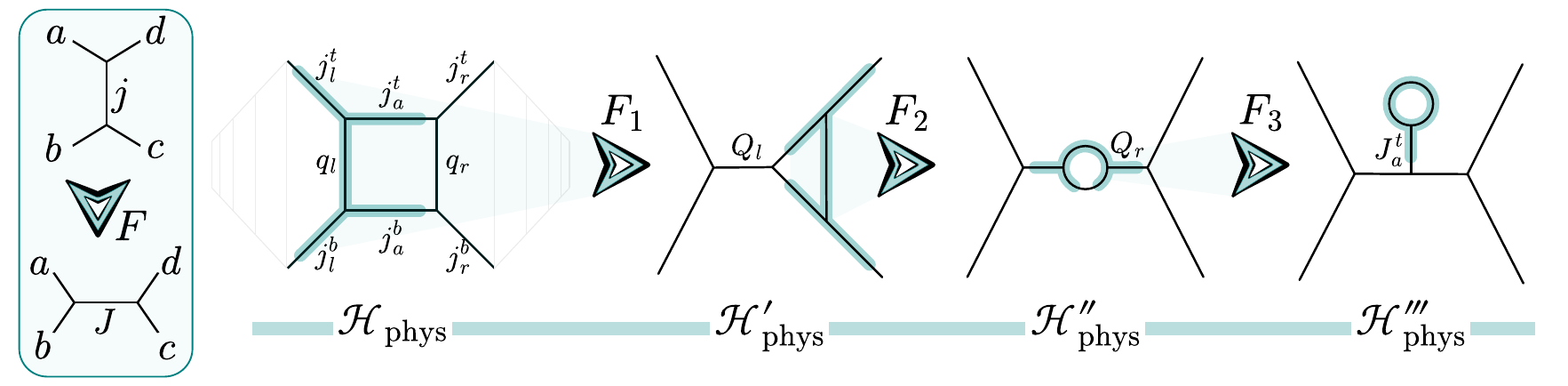}
    \caption{An F move takes an original two-vertex diagram (with vertices $\{a,d,j\}$ and $\{b,c,j\}$) to a modified diagram (with vertices $\{a,b,J\}$ and $\{d,c,J\}$) with updated $j \rightarrow J$ as shown in the box on the left. Links $a,b,c$ and $d$ are controls and remain unchanged. The matrix element for this transition is governed by Eq.~\eqref{eq: F-move} and~\eqref{eq: q-deformed F matrix}. Active links of the plaquette on the original lattice (far left) are $|q_l\rangle, |j_a^t\rangle, |j_a^b\rangle, |q_r\rangle$. Links $|j_l^t\rangle, |j_l^b\rangle, |j_r^t\rangle, |j_r^b\rangle$ are external but connected to the plaquette at its vertices. These links serve as controls in the reduced basis with projection degrees of freedom integrated out. Each step in the F-sequence turns the active link it acts on into an external control link (indicated by capitalized link labels). The sequence of F-moves shown diagrammatically here alters the contractions of lattice vertices so that the diagonalized plaquette operator acts only on the $|j_a^b\rangle$ register.}
    \label{fig: f sequence}
\end{figure*}

When evolving physical (gauge-invariant) states in the absence of matter, a gauge singlet will continue to be present at each vertex throughout time evolution. Employing a component of this symmetry allows integration of the projection quantum numbers ($m_l, m_l'$)~\cite{Banuls:2017ena,Klco:2019evd,ARahman:2021ktn,Ciavarella:2021nmj,Ciavarella:2022zhe}. In this reduced basis, the wave function for a 1D $L$-plaquette chain with PBCs can be written as a superposition of product states of each link's irrep $|j_1j_2...j_{3L}\rangle$, and the plaquette operator is controlled by external links present at the four vertices. Matrix elements of the plaquette operator can be written in terms of Wigner $6j$ symbols as provided in Ref.~\cite{Klco:2019evd}, whose notation we have adopted in Fig.~\ref{fig: f sequence}. For our purposes, it is convenient to rewrite these elements in terms of F-symbols as 
\begin{multline}\label{eq: plaquette F-symbols}
\langle ...q_l' j_a^{t \prime} j_a^{b \prime} q_r'... | \Box | ...q_l j_a^t j_a^b q_r... \rangle \\ = (-1)^{-\Delta j_a^t - \Delta j_a^b + \Delta q_l + \Delta q_r}
\times\\
\quad \quad 
\begin{bmatrix} 
      j_{l}^{t} & j_{a}^{t} & q_{l} \\ 
      1/2 & q_{l}^{\prime} & j_{a}^{t\prime}
\end{bmatrix}
\begin{bmatrix} 
      j_{l}^{b} & q_{l} & j_{a}^{b} \\ 
      1/2 & j_{a}^{b\prime} & q_{l}'
\end{bmatrix}
\times\\
\qquad 
\begin{bmatrix} 
      j_{r}^{t} & q_{r} & j_{a}^{t} \\ 
      1/2 & j_{a}^{t\prime} & q_{r}'
\end{bmatrix}
\begin{bmatrix} 
      j_{r}^{b} & j_{a}^{b} & q_{r} \\ 
      1/2 & q_{r}^{\prime} & j_{a}^{b\prime}
\end{bmatrix} \ \ \ , 
\end{multline}
where $\Delta j = j' - j$, $D(j) = 2j+1$ is the dimension of irrep $j$, and F-symbols are defined in terms of Wigner $6j$'s in the form
\begin{equation} 
\begin{gathered}
    F^{abe}_{cdf} := 
    \begin{bmatrix}
        a & b & e \\ 
      c & d & f
    \end{bmatrix}
    \\
    =
    (-1)^{a+b+c+d}\sqrt{D(e)D(f)} 
    \begin{Bmatrix} 
      a & b & e \\ 
      c & d & f
    \end{Bmatrix} \ \ \ .
    \end{gathered}
    \label{eq: F matrix}
\end{equation}
Note that $\Delta j$ is always a half-integer, and thus the exponent is always an integer. 
Flux values not included in the bra-ket are unchanged by the operator and bear an implied $\delta_{jj'}$, i.e., external links ($|j_l^t\rangle, |j_l^b\rangle, |j_r^t\rangle,$ and $|j_r^b\rangle$) are control registers that affect the matrix element, but are unaffected by the action of the operator.

\section{The role of q-deformation}\label{sec: Main: the role of q-deformation}

Circuit synthesis of the plaquette operator for general angular momentum truncations can be cumbersome. To alleviate such costs, Ref.~\cite{Zache:2023dko} proposes to use a sequence of \textit{F-moves} that partially diagonalize the plaquette operator by altering vertex contractions on the lattice. Such F-moves, proposed in the 1980s as a way to move between the electric and magnetic (plaquette) eigenbases~\cite{Robson1982}, are unitary in the physical subspace of untruncated lattice Yang-Mills theories. However, this unitarity is \textit{not} generally retained upon local flux truncation.
By q-deforming the gauge field, these features can be achieved simultaneously, i.e., the diagonalization procedure can be made unitary in a finite-dimensional physical subspace~\cite{Zache:2023dko}, allowing implementation on a finite quantum device.

\subsection{The F-move sequence}\label{sec: Main: non-deformed F-moves}
The role of \textit{F-moves}, originally formulated for the untruncated lattice~\cite{Robson1982}, is to rearrange angular momentum contractions within two-vertex diagrams, as shown in the left panel of Fig.~\ref{fig: f sequence}. As operators, their matrix elements are governed by the F-symbols of Eq.~\eqref{eq: F matrix},
\begin{equation} \label{eq: F-move}
    \langle abcdJ| F | abcdj\rangle = 
    \begin{bmatrix}
        a&b&J\\
        c&d&j
    \end{bmatrix} \ \ \ .
\end{equation}
These F-moves constitute a unitary transformation over the space of gauge-invariant states with untruncated local angular momentum. Initial states with physical flux (satisfying Gauss's law) on the original lattice are unitarily mapped to states with physical flux on the modified lattice (i.e., $F: \mathcal{H}_{\text{phys}} \rightarrow \mathcal{H}_{\text{phys}}'$), while unphysical states are annihilated by the F-move.

On a section of the lattice containing a single plaquette and its neighboring external controls, we perform a series of F-move contractions in order to shrink the active space (closed loop) of the plaquette operator, as shown in Fig.~\ref{fig: f sequence}. On the modified lattice with vertex connections changed by the F-sequence, the transformed $|q_l\rangle, |q_r\rangle$ and $|j_a^t \rangle$ registers join the external links as controls, and the plaquette operator acts non-trivially only on the closed loop of the $|j_a^b\rangle$ register. A final G-move (not visualized here) diagonalizes the plaquette operator on the $|j_a^b\rangle$ register, completing the transformation to the magnetic basis \cite{Robson1982}.

\subsection{Q-deformed F-moves on the truncated lattice}\label{sec: q-deformed F-moves}
In order to render F-moves unitary over the locally truncated Hilbert space amenable to quantum computational devices, we q-deform the symmetry group of the theory following Ref.~\cite{Zache:2023dko}. The quantum-group description involves deforming the generators of the Lie algebra, tuned by a \textit{deformation} parameter $q$ such that the original Lie group is recovered when $q = 1$. In order to ensure finite-dimensional irreps (as desired for our truncated simulation) we choose that $q$ is a root of unity (see, for example, Ch.~2 of Ref.~\cite{Biedenharn1995} and Ref.~\cite{Kirillov:1991ec}),
\begin{equation}\label{eq: deformation parameter}
    q = e^{i\frac{2\pi}{k+2}} \ \ \ ,
\end{equation}
where $q$ is parametrized by $k$, an integer which here corresponds to local flux truncation $k/2$. This choice of deformation parameter ensures closure of the q-deformed group under truncation, which is essential to ensure that the q-deformed F-moves are unitary in the truncated gauge-invariant subspace. For every $n \in \mathbb{R}$, a q-number is subsequently defined as
\begin{equation}\label{eq: q-number}
    [n]_k = \frac{\text{sin}(\frac{\pi n}{k + 2})}{\text{sin}(\frac{\pi}{k + 2})} \ \ \ .
\end{equation}
The procedure of q-deformation promotes values in the original description of the Lie algebra to q-numbers. For example, in SU(2)$_k$, the commutation relation $[J_+^{(k)}, J_-^{(k)}] = 2J_z^{(k)}$ persists, where the eigenvalues of the operator on the RHS are now q-numbers~\cite{Biedenharn1995}. 
From Eq.~\eqref{eq: F matrix},  q-deformed F-symbols are constructed by promoting integers in the irrep dimension $D(j)$ and Racah formula expansion of the $6j$ symbols to q-numbers,
\begin{multline} \label{eq: q-deformed F matrix}
    \begin{bmatrix}
        a&b&e\\
        c&d&f
    \end{bmatrix}_k = (-1)^{a+b+c+d}\\\sqrt{D_k(e)D_k(f)} 
    \begin{Bmatrix} 
      a & b & e \\ 
      c & d & f
    \end{Bmatrix}_{k} \ \ \ ,
\end{multline}
where the subscript on the curly brackets indicates the q-deformed $6j$ symbol and $D_k(j)= [2j+1]_k$ is the q-deformed quantum dimension. Going forward, we will refer only to q-deformed F-symbols, so we will usually suppress this subscript notation. Further details in the definition of q-deformed F-symbols are provided in Appendix~\ref{sec: Appendix: F definition}. 

The replacement of F-symbols in Eq.~\eqref{eq: plaquette F-symbols} by Eq.~\eqref{eq: q-deformed F matrix} constitutes the q-deformation of the plaquette operator for the Hamiltonian of SU(2)$_k$ Yang-Mills.
Note that the Casimir invariant appearing in the electric term of the Hamiltonian, $E^2$, may also be q-deformed such that $E^2_k|j\rangle = [j]_k[j+1]_k|j\rangle$ \cite{Biedenharn1995}. 
Because the convergence is so far observed to be insensitive to the electric deformation, our circuit strategy focuses solely on magnetic deformation.

Under q-deformation, F-symbols (and their antecedent Wigner $6j$ symbols) retain their original symmetries and orthogonality relations. However, q-deformation inheres an additional constraint. In the original (non-deformed) representation, the $6j$ symbols encoded in the magnetic term of the Eq.~\eqref{eq: plaquette F-symbols} Hamiltonian enforce gauge invariance by requiring that flux at each vertex ($j_1,j_2,j_3$) has an integer sum and satisfies the triangle inequalities,
\begin{align}\label{eq: gauge singlet non-def}
 j_1 + j_2 + j_3 &\in \mathbb{N}_0 \ \ , \nonumber\\
    j_1 + j_2 \geq j_3\ \ ,  \ \  j_2 + j_3 &\geq  j_1 \ \ , \ \  j_3 + j_1 \geq j_2 \ \ , 
\end{align}
where $\mathbb{N}_0$ is the set of non-negative integers. In general, for local truncation $k/2$, the sum of angular momenta at each vertex is maximally $3k/2$. However, due to our choice of deformation parameter as a root of unity, a stricter condition, the \textit{fusion constraint},
\begin{equation}\label{eq: fusion constraint}
    j_1 + j_2 + j_3 \leq k \ \ \ ,
\end{equation} 
is satisfied at each vertex throughout time evolution, defining a finite-dimensional physical subspace in which q-deformed F-moves are unitary.
These gauge-invariant constraints (including the fusion constraint) will be packaged into a vertex admissibility function, $\delta_{j_1j_2j_3}$, which is enforced by the q-deformed $6j$ symbol (Eq.~\eqref{eq: Racah 6j}).
The q-deformation therefore restricts the physical Hilbert space of the q-deformed theory $\mathcal{H}_\text{phys}^k(k)$ to be a subspace of the non-deformed gauge-invariant Hilbert space $\mathcal{H}_\text{phys}(k)$ subject to the same local truncation, $d = k+1$. 
The gauge-invariant states that exceed the fusion constraint are those with high local energy density, such that the q-deformed theory experiences both a link-local and a vertex-local truncation softening the concentration of flux at the UV scale of the lattice.

\subsection{Remarks on convergence with raised truncation}\label{sec: Main: convergence}
The present q-deformation imposes an approximation parameter $k$ to the group structure that also serves as the local field truncation.  The non-deformed lattice theory is recovered with increasing local Hilbert space dimension in the limit $k \rightarrow \infty$ (where $q = 1$). 
Explorations as a function of $k$ and coupling $g^2$ report promising convergence properties~\cite{Zache:2023dko, Hayata:2026xeo}. 
For example, in Ref.~\cite{Zache:2023dko}, ground state observables of the q-deformed theory on an infinite lattice of plaquettes were found to converge rapidly for $k$ above a critical value $k_c \simeq 4.4/g - 2.5$. 
The subsequent expectation of in-practice manageable local Hilbert space dimensions is consistent with neighboring explorations of field quantum simulation~\cite{Jordan:2011ci,Jordan:2011ci,Klco:2018zqz,Davoudi:2020yln,Tong:2021rfv,Ciavarella:2025tdl}.
In modest time-dependent calculations, we observe similarly rapid convergence of low-energy dynamical observables.  

\begin{figure}
    \centering
    \includegraphics[width=0.98\linewidth]{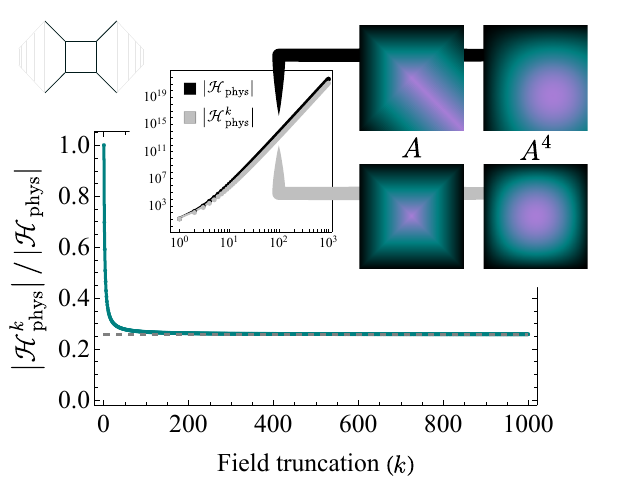}
    \caption{Size of q-deformed physical subspace (satisfying gauge invariance and fusion constraint) relative to that of the non-deformed physical subspace, both truncated at local Hilbert space dimension $d = k+1$, for an eight-link lattice section spanned by a plaquette operator. 
    An order-8 polynomial fit applied to $|\mathcal{H}_{\text{phys}}|$ estimates that the ratio converges to the dashed line at $0.2563(5)$}.
    \label{fig: state retention}
\end{figure}
To gain further insight on the rate of convergence that can be anticipated for the q-deformed theory, we quantify the impact of the fusion constraint on the dimension $|\mathcal{H}_\text{phys}^k|$ of a plaquette operator in Fig.~\ref{fig: state retention}.
The presented fusion constraint retention ratio   $|\mathcal{H}_\text{phys}^k|/|\mathcal{H}_\text{phys}|$ informs the relative rates of convergence of the truncated theories (q-deformed and non-deformed) to the infinite physical Hilbert space of the untruncated theory.
For the lowest truncations, very few physical states in the non-deformed theory violate the fusion constraint. 
For higher truncations, however, a significant portion of states in $\mathcal{H}_\text{phys}$ are excluded from $\mathcal{H}_\text{phys}^k$. 
Encouragingly however, the fusion constraint retention ratio $|\mathcal{H}_\text{phys}^k|/|\mathcal{H}_\text{phys}|$ is observed to stabilize to a fixed value at large $k$ calculated to be $0.2563(5)$, where the error bar expresses an estimate of systematics.  We therefore expect that the q-deformed theory converges to the untruncated theory with comparable scaling to the non-deformed truncated theory.

To create Fig.~\ref{fig: state retention}, we leverage transfer matrix techniques, for example as in Section~III of Ref.~\cite{Pato:2026wow}, in order to (1) calculate the physical Hilbert space dimension of an 8-link plaquette operator at truncations well beyond those whose bases can be classically enumerated and (2) provide an analytic expression for the physical dimension $|\mathcal{H}_{\text{phys}}^k|$ governing the plaquette operator in the q-deformed theory.

Consider the space of the plaquette operator as a length-4 periodic contraction of vertex elements. 
Construct the vertex adjacency matrix $A$ such that each matrix element $A_{\ell_1\ell_2}$ is the multiplicity of external link irreps given active plaquette links of values $\ell_1/2$ and $\ell_2/2$.
For example, for a 4-dimensional local field truncation ($k = 3$), the $4\times4$ adjacency matrix has a final element of $A_{3,3} = 2$ in the non-deformed theory counting configurations with external flux $j = 0, 1$, while $A^k_{3,3} = 1$ in the q-deformed theory as the $j =1$ configuration is eliminated via the fusion constraint. 
Generically, these matrix elements are
\begin{multline}
A_{\ell_1\ell_2} = \frac{1}{2} 
\Big( \min\left[ \ell_1+\ell_2, k-(\ell_1+\ell_2+k)_{\text{mod\ }  2}\right]  \\ -  |\ell_1-\ell_2| \Big)+1 \ \ \ ,
\label{eq:adjMxEl}
\end{multline}
\begin{multline}
A^k_{\ell_1\ell_2} = \frac{1}{2}\Big(\min\left[ \ell_1 + \ell_2, 2k - \ell_1-\ell_2\right]  \\ - |\ell_1-\ell_2|\Big)+1  \ \ \ 
\label{eq:adjMxElqdef}
\end{multline}
for the non-deformed and q-deformed theories, respectively. The first term captures the integer index of the maximum possible flux on the external link, where the $\left(\text{mod\ } 2\right)$ shift in Eq.~\eqref{eq:adjMxEl} assures that the maximum has the same parity as~$(\ell_1+\ell_2)$ when restricted by the local field truncation.
After the difference captures the span from maximum to minimum valid flux indices, every other index is counted, i.e., all integer or half-integer irreps according to the parity of~($\ell_1+\ell_2$).
Once this $d \times d$ adjacency matrix of multiplicities is calculated, the physical subspace dimension can be expressed for either theory as
\begin{equation}
\left|\mathcal{H}_{\text{phys}}^{(k)}\right| = \text{Tr}\left[ \left(A^{(k)}\right)^4\right] \ \ \ .
\label{eq:TrA4}
\end{equation}
These dimensions and their ratio are shown in the inset and main panel of Fig.~\ref{fig: state retention}. In order to observe their relative scaling, truncations significantly larger than anticipated to be needed in practical quantum simulations are included.

As will be demonstrated several times in this work, the additional structure furnished by the q-deformation leads to improved calculability. 
Illustrated for $k = 100$ at the right of Fig.~\ref{fig: state retention}, the adjacency matrix for the q-deformed theory is not only symmetric but has $D_4$ symmetry and can be written as a simple sum of rank-one components as
\begin{equation}
A^k =
\sum_{\ell=1}^{k+1} 
|\ell\rangle_k \langle \ell|_k \ \delta_{(k+1)\text{mod\ } 2,\ \ell\ \text{mod\ }2} \ \ \ ,   
\end{equation}
where 
$|\ell\rangle_k = \begin{pmatrix}
0 & 
\cdots & 
0 & 
1 &
\cdots &
1 &
0 & 
\cdots &
0
\end{pmatrix}^T$ is a $(k+1)$-dimensional vector with a centered domain of $\ell$ unit elements and the Kronecker delta enforces a parity constraint between $k+1$ and $\ell$, i.e., the sum increments in steps of two and begins one value higher if $k$ is odd.
Because $\langle \ell_1|\ell_2\rangle =\min\left[\ell_1, \ell_2\right]$, the trace of the fourth power in Eq.~\eqref{eq:TrA4} can be written succinctly as
\begin{multline}
\left|\mathcal{H}_{\text{phys}}^{k}\right| = \sum_{\ell_1, \ell_2, \ell_3, \ell_4=1}^{k+1} 
\min\left[\ell_1, \ell_2\right] 
\min\left[\ell_2, \ell_3\right]\times \\
\qquad \qquad \min\left[\ell_3, \ell_4\right]
\min\left[\ell_4,\ell_1\right] \times \\  \prod_{i = 1}^4 \delta_{(k+1)\text{mod\ } 2,\ \ell_i\ \text{mod\ }2} \ \ \ .
\end{multline}
Evaluating this sum yields separate expressions for odd and even $k$ truncation,
\begin{equation}
    \left|\mathcal{H}_{\text{phys}}^{k }\right| = \begin{cases}
       h(k)  & k \text{ odd} \\
       h(k)+\frac{315}{10080}  & k \text{ even}
    \end{cases} \ \ \ ,
\end{equation}
with
\begin{multline}
h(k) =  \frac{1}{10080}\left(1152 d + 2048 d^2 + 2464 d^3  \right. \\  + 2128 d^4 + 1288 d^5 + 532 d^6  \\ \left.+ 
   136 d^7 + 17 d^8\right)  
\end{multline}
and $d = k+1$. 
By exact analysis, the order-8 polynomial scaling of the physical Hilbert space relevant to the plaquette operator is retained in the presence of the fusion constraint when q-deforming the gauge field.
This is further evidence that q-deformation does not significantly compromise the approach to continuum field values.

\section{F-sequence circuitry via gauge-variant completions}\label{sec: Main: circuitry}
So far, q-deformation has provided a path for diagonalizing the plaquette operator in the presence of local flux truncations by enabling q-deformed F-moves that are unitary within $\mathcal{H}_{\text{phys}}^k$.  
While such progress is crucial, this unitarity is insufficient to furnish a full quantum simulation strategy. As is the case for F-moves generically, q-deformed F-moves continue to annihilate unphysical states, and are thus non-unitary in the unphysical Hilbert space, i.e.,  outside $\mathcal{H}_{\text{phys}}^k$. 

In order to realize a diagonalizing F-move on a quantum device whose computational Hilbert space contains gauge-variant or fusion-violating states as well, an operator completion must be employed to produce unitary gates. 
As discussed in \cite{Klco:2019evd}, the isolation of dynamics to the physical subspace allows any gauge-variant completion (GVC) to be selected that may be desirable for device implementation. 
That is, when interested specifically in the behavior of physical states, errors that arise within the unphysical subspace may be subjected to independent evolution chosen to support circuit synthesis.

In Section~\ref{sec: qdefDiagF1}, GVC freedom will be leveraged to identify and construct a series of F-moves that systematically reduce the spatial extent of the plaquette operator throughout the diagonalization procedure.
In Section~\ref{sec:circuitStrategy}, a second stage of the GVC will be used to synthesize associated unitary circuits.

\subsection{Phased diagonalization and operator compression}
\label{sec: qdefDiagF1}

Following the diagrammatic procedure of lattice contractions~\cite{Robson1982, Zache:2023dko} (shown in Fig.~\ref{fig: f sequence}), we identify the following “phased" F-moves
\begin{subequations}\label{eq: phased F moves}
    \begin{equation}\label{eq: F1 def}
    \langle ...Q_l...|F_1|...q_l...\rangle = (-1)^{-q_l - Q_l}
    \begin{bmatrix}
        j_l^t & j_l^b & Q_l\\
        j_a^b & j_a^t & q_l
    \end{bmatrix}
    \end{equation}
    \begin{equation}\label{eq: F2 def}
    \langle ...Q_r...| F_2| ...q_r...\rangle = (-1)^{-q_r - Q_r}
    \begin{bmatrix}
        j_a^t & j_a^b & Q_r\\
        j_r^b & j_r^t & q_r
    \end{bmatrix}
    \end{equation}
    \begin{equation}\label{eq: F3 def}
    \langle ... J_a^{t}...|F_3 |... j_a^t ... \rangle  = (-1)^{j_a^t + J_a^{t}}
    \begin{bmatrix}
        q_l & q_r & J_a^{t}\\
        j_a^b & j_a^b & j_a^t
    \end{bmatrix}\ \ \ ,
    \end{equation}  
\end{subequations}
where the phases that multiply the F-symbols are carefully chosen in order to cancel with those in Eq.~\eqref{eq: plaquette F-symbols}. Thus, active registers reside in the right-most column of the F-symbol, while control registers are those in the left and center columns.

As an example, consider evaluation of $F_1$, which diagonalizes the plaquette operator over the $|q_l\rangle$ register.   The protocol of Fig.~\ref{fig: f sequence} identifies the control and active spaces for $F_1$, leading to an ansatz of the form 
\begin{equation}
    \langle ...Q_l...|F_1|...q_l...\rangle = (-1)^{\alpha_1} \begin{bmatrix}
        j_l^t & j_l^b & Q_l\\
        j_a^b & j_a^t & q_l
    \end{bmatrix} \ \ \ . \nonumber
\end{equation}
The plaquette operator after the action of $F_1$ is 
\begin{multline}
    \langle ...Q_l' j_a^{t \prime} j_a^{b \prime} q_r' ...| F_1 \Box F_1^\dagger | ... Q_l j_a^{t} j_a^{b} q_r  ...\rangle \\
    = \sum_{q_l, q_l'} (-1)^{\gamma + \alpha_1' - \alpha_1}
    \begin{bmatrix} 
          j_{l}^{t} & j_{l}^{b} & Q_l' \\ 
          j_a^{b\prime} & j_a^{t\prime} & q_l'
    \end{bmatrix}
    \begin{bmatrix} 
          j_{l}^{t} & j_{a}^{t} & q_l \\ 
          1/2 & q_l' & j_{a}^{t\prime}
    \end{bmatrix} 
    \\
    \times
    \begin{bmatrix} 
          j_{l}^{b} & q_l & j_{a}^{b} \\ 
          1/2 & j_{a}^{b\prime} & q_l'
    \end{bmatrix}
    \begin{bmatrix} 
          j_{r}^{t} & q_{r} & j_{a}^{t} \\ 
          1/2 & j_{a}^{t\prime} & q_{r}^{\prime}
    \end{bmatrix}
    \\
    \times
    \begin{bmatrix} 
          j_{r}^{b} & j_{a}^{b} & q_{r} \\ 
          1/2 & q_{r}^{\prime} & j_{a}^{b\prime}
    \end{bmatrix} 
    \begin{bmatrix}
       j_l^t & j_l^b & Q_l \\
       j_a^b & j_a^t & q_l
    \end{bmatrix}  \ \ \ ,
\end{multline}
where $\gamma = -\Delta j_a^t - \Delta j_a^b + \Delta q_l + \Delta q_r$ is the exponent in Eq.~\eqref{eq: plaquette F-symbols} and capitalized indices correspond to quantum registers on the modified lattice. By choosing $\alpha_1' = -Q_l' - q_l'$ and $\alpha_1 = -q_l - Q_l$ as in Eq.~\eqref{eq: F1 def}, the phase becomes $\gamma + \alpha_1' - \alpha_1 = -\Delta j_a^t - \Delta j_a^b - \Delta Q_l + \Delta q_r$, which is now independent of the summed indices $q_l, q_l'$. This allows application of the standard F-symbol pentagon identity (Eq.~\eqref{eq: F pentagon identity}) and orthogonality relation (Eq.~\eqref{eq: F orthogonality relation}), utilizing symmetry properties detailed in Appendix \ref{sec: Appendix: properties} as needed. The pentagon identity analytically performs the sum over $q_l'$
\begin{equation}
\begin{gathered}
\sum_{q_l'}
\begin{bmatrix} 
      j_{l}^{t} & j_{l}^{b} & Q_l' \\ 
      j_a^{b\prime} & j_a^{t\prime} & q_l'
\end{bmatrix}
\begin{bmatrix} 
      j_{l}^{t} & j_{a}^{t} & q_l \\ 
      1/2 & q_l' & j_{a}^{t\prime}
\end{bmatrix}
\begin{bmatrix} 
      j_{l}^{b} & q_l & j_{a}^{b} \\ 
      1/2 & j_{a}^{b\prime} & q_l'
\end{bmatrix}
\\
= 
\begin{bmatrix} 
      1/2 & j_a^{b \prime} & j_a^b \\
      Q_l' & j_a^t & j_a^{t \prime}
\end{bmatrix}
\begin{bmatrix} 
      j_a^t & j_l^t & q_l \\
      j_l^b & j_a^b & Q_l'
\end{bmatrix} \ \ \ ,
\end{gathered}
\end{equation}
by identifying $\mathbf{j} = \{ 1/2, j_a^{b\prime}, j_l^b, q_l, j_a^b, j_a^t, j_l^t, j_a^{t\prime}, Q_l'\}$ and $J = q_l'$ from the notation of Eq.~\eqref{eq: F pentagon identity}. The remaining sum over $q_l$ may then be addressed by the orthogonality relation
\begin{equation} 
    \sum_{q_l} 
    \begin{bmatrix} 
          j_a^t & j_l^t & q_l \\
          j_l^b & j_a^b & Q_l'
    \end{bmatrix}
    \begin{bmatrix}
       j_l^t & j_l^b & Q_l \\
       j_a^b & j_a^t & q_l
    \end{bmatrix}
    = \delta_{Q_l, Q_l'} \ \ \ .
\end{equation}
Combining results, this first step in the
plaquette operator diagonalization leaves the $|q_l\rangle$ register subsequently stationary and reduces the number of F-symbols characterizing the matrix element by one,
\begin{equation}
\begin{gathered}
    \langle ... Q_l' j_a^{t \prime} j_a^{b \prime} q_r'  ...| F_1 \Box F_1^\dagger | ... Q_l j_a^{t} j_a^{b} q_r ...\rangle \\
    = (-1)^{-\Delta j_a^t - \Delta j_a^b - \Delta Q_l + \Delta q_r} \delta_{Q_l, Q_l'} \textcolor{Purple!90!Black}{\delta_{j_l^t j_l^b Q_l}}
    \\
    \times
    \begin{bmatrix} 
        1/2 & j_a^{b \prime} & j_a^b \\ 
        Q_l' & j_a^t & j_a^{t \prime}
    \end{bmatrix}
    \begin{bmatrix} 
          j_{r}^{t} & q_{r} & j_{a}^{t} \\ 
          1/2 & j_{a}^{t\prime} & q_{r}^{\prime}
    \end{bmatrix}
    \begin{bmatrix} 
          j_{r}^{b} & j_{a}^{b} & q_{r} \\ 
          1/2 & q_{r}^{\prime} & j_{a}^{b\prime}
    \end{bmatrix}  \ \ \ ,
\end{gathered}
\label{eq:f1boxf1dag}
\end{equation}
where $\Delta Q_l = 0$ due to $\delta_{Q_l, Q_l'}$. 

Notice that in the right side of Eq.~\eqref{eq:f1boxf1dag} we have inserted a vertex admissibility function $\delta_{j_l^t j_l^b Q_l}$ (highlighted in purple) that enforces the Gauss's law and fusion constraints over the left-most vertex in the second diagram of Fig.~\ref{fig: f sequence}. 
This insertion allows the equality of Eq.~\eqref{eq:f1boxf1dag} to hold in the computational Hilbert space beyond the gauge-invariant subspace.
Furthermore, the $\delta_{Q_l,Q_l'}$ that arises after $F_1$ 
causes this vertex to remain fixed throughout the remaining procedure. 
Thus, violations in the physicality of this vertex will not mix into the gauge-invariant subspace, making it possible to choose a GVC that is agnostic to the physicality of this vertex. 
In circuit language, enforcing the admissibility function $\delta_{j_l^t j_l^b Q_l}$ would require a non-unitary controlled operator to check Gauss's law and the fusion constraint at that vertex.  
However, we can select a GVC that erases this control (removes  $\delta_{j_l^t j_l^b Q_l}$) without impact to the physical evolution.
Such a choice of GVC following $F_1$ allows the subsequent plaquette operator circuit to act over a reduced portion of the lattice. In turn, this allows the diagonalized plaquette operator time evolution to be reduced to a single-qudit subspace with a single-qudit control register, central to our aims in this work. 

Mathematical details of the diagonalization associated with the remaining phased F-moves are provided in Appendix~\ref{sec: Appendix: diagonalization}.
Following the sequence of phased F-moves, the matrix elements of the transformed plaquette operator $\Box''' = F_3F_2F_1\Box F_1^\dagger F_2^\dagger F_3^\dagger$ are
\begin{equation}\label{eq: plaq after F3}
 \langle ...j_a^{b\prime}...|\Box'''|... j_a^b...\rangle
 = (-1)^{-\Delta j_a^b} \begin{bmatrix} 
      j_a^b & 1/2 & j_a^{b \prime} \\
      j_a^{b \prime} & J_a^t & j_a^b
\end{bmatrix},
\end{equation}
where we have chosen a GVC that eliminates the three additional vertex admissibility functions that arise in the diagonalization. As desired, only the $|j_a^t\rangle$ register now acts as a control to the $|j_a^b\rangle$ remaining active register. 
In order to complete the diagonalization, unitary matrices $G(J_a^t)$ can be identified for each control sector 
\begin{equation}
    \Box''' = \prod_{J_a^t}G^\dagger(J_a^t)\tilde{\Box}(J_a^t)G (J_a^t)  \ \ \ ,
\end{equation}
which diagonalize the action of the plaquette operator over the $|j_a^b\rangle$ register.

\subsection{Flux hierarchy inversion symmetry}\label{sec: Main: FHI}

Section~\ref{sec: q-deformed F-moves} discusses how the fusion constraint eliminates states of high flux density from the physical subspace of the q-deformed theory. 
As such, the fusion constraint can be understood as an extension of the link truncation of high-flux irreps to a truncation of concentrated flux at vertices. 
Logically, low-energy regimes of the q-deformed theory continue to converge more rapidly than their high-energy counterparts as the untruncated theory is approached~\cite{Zache:2023dko}.
However, it is important to note that q-deformation impacts interactions at all scales of the theory, not only near the UV boundary. 

While this impact can be seen directly as Eq.~\eqref{eq: q-deformed F matrix} q-deforms the vertex factors in Eq.~\eqref{eq: plaquette F-symbols}, there are structural properties that illuminate not only the impact but the opportunities afforded by q-deformation of the group structure.
For example, Section~\ref{sec: Main: convergence} shows that the vertex adjacency matrix used to calculate the physical dimension relevant to the plaquette operator gains $D_4$ symmetry upon q-deformation.  
Specifically, the gained persymmetry component of this $D_4$ over external link multiplicities forecasts the presence of a symmetry relating low- and high-energy scales.
At the level of interactions beyond multiplicities, it is found that persymmetry is also manifest in $\Box'''$.
Physically, this is a \textit{flux hierarchy inversion} symmetry, in which $\langle j'|\Box'''|j\rangle$ is invariant upon replacement of $j, j'$ with corresponding irreps reflected across the center of the angular momentum range $\{0, ...k/2\}$.
Proving the presence of this symmetry is detailed in Appendix~\ref{Appendix: FHI symmetry}.

Although we do not leverage it for circuit optimizations in the present calculations, this symmetry halves the number of matrix elements of $\Box'''$ to be classically calculated. 
In Section~\ref{sec: Main: resource scaling}, we describe other structural features of $\Box'''$ and the phased F-moves that we leverage to reduce quantum circuitry resource requirements.

\subsection{Construction of F-move unitaries}
\label{sec:circuitStrategy}

For our final step of determining unitary operators capturing the diagonalization procedure in the physical subspace, each phased F-move can be characterized by a product of controlled single-qudit unitaries, where a GVC is imposed over the gauge-variant subspace otherwise annihilated by $F$,
\begin{equation}
    U_F = \prod_{\text{controls}} U_F(a,b,c,d)  \ \ \ .
\end{equation}
In order to identify an appropriate series of such unitaries, we take a diagrammatic approach. 
Considering a single two-vertex, five-link diagram (left panel of Fig.~\ref{fig: f sequence}), each phased F-move acts non-trivially only between initial(final) states with physical flux satisfying the q-deformed gauge singlet conditions on the original(modified lattice). 
For each physical control sector, we (1) identify valid flux configurations of the active link on the original and modified lattices (2) assign to these matrix elements the appropriate transition amplitudes determined by the phased F-move of Eq.~\eqref{eq: phased F moves} and (3) determine a GVC that naturally completes the unitary in the active Hilbert space. 

\begin{figure}
    \centering
    \includegraphics[width=0.95\linewidth]{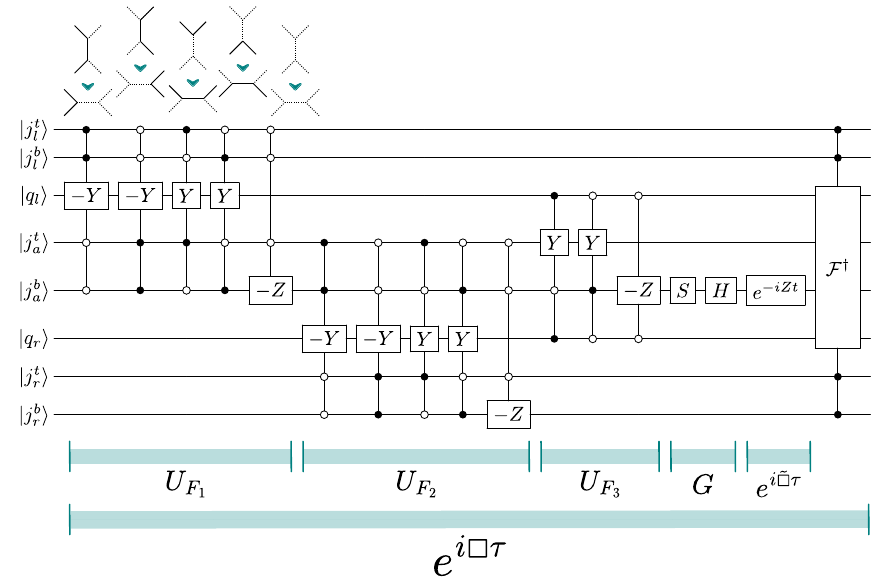}
    \caption{Circuit capable of implementing plaquette operator time evolution ($\tau = \frac{t}{g^2N_T}$ for $N_T$ Trotter steps) via F-sequence diagonalization, demonstrated for the qubit-truncated gauge field ($k=1$). 
    By rotating from the eigenbasis of the electric operator to that of the compressed plaquette operator, the time evolution is isolated to a single-qudit active space on the modified lattice ($|j_a^b\rangle$).
    Each step in the diagonalization can be accomplished by a series of multi-controlled single qudit gates. 
    Each control sector corresponds to a set of external links that allow $\mathcal{H}_\text{phys} \rightarrow \mathcal{H}_\text{phys}'$ transitions, as indicated by the diagrams above $U_{F_1}$ components.
    }
    \label{fig: schematic circuit}
\end{figure}
For example, consider the third flux diagram above the $U_{F_1}$ circuit in Fig.~\ref{fig: schematic circuit}. 
The controls $2\mathbf{j} = \{ 1, 0, 0, 1\}$ admit only an active link of $2q_l = 0$ on the original lattice and only an active link of $2Q_l = 1$ on the modified lattice, depicted by dashed and solid lines respectively.
Thus, the flux through the active link must change in order to maintain physicality, $F: \mathcal{H_{\text{phys}}} \rightarrow \mathcal{H'_{\text{phys}}}$.
Calculating the $\langle Q_l=1/2 | U_{F_1} |q_l=0\rangle$ matrix element from Eq.~\eqref{eq: F1 def} leads to the following unitary, 
\begin{equation} U_{F_1}(1,0,0,1) = 
\begin{bmatrix}
   0 & * \\
   -i & 0
\end{bmatrix} \ \ \ ,
\end{equation}
where $* = i$ is a GVC chosen such that $U_{F_1}(1,0,0,1) = -Y$.
Proceeding in this way for each control sector $\{a,b,c,d\}$, we identify an appropriate series of gates for each F-move by considering allowed active link transitions that maintain gauge invariance as the vertex contractions are altered.
These techniques apply for any $k$ truncation, detailed in Appendix~\ref{sec: Appendix: recipe}.  
A full account of such circuit elements is provided for the qubit $k = 1$  and qutrit $k = 2$ truncations in Appendix~\ref{Appendix: circuits}.

In crafting the quantum circuit of Fig.~\ref{fig: schematic circuit} from the unitaries determined through this process, two further simplifications have been made,  representing steps that continue to be available at higher $k$ truncations.  
First, though there are in total eight diagrams capturing valid transitions to the modified lattice, only five---those depicted above $U_{F_1}$---contribute to the quantum circuitry. Four of the eight transitions are characterized by diagonal $\mathcal{H}_{\text{phys}} \rightarrow \mathcal{H}_{\text{phys}}'$ matrix elements (see Table~\ref{tab:quditF1F2F3Gunitaries}).  
By selecting an overall -1 phase on all operators (e.g., $-Y \rightarrow Y$ for  $U_{F_1}(1, 0, 0, 1)$), three of these four can be GVC completed as the identity $\mathbb{I}$ and need not contribute circuit elements.
The fourth, $U_{F_1}(0,0,0,0)$, can then be completed as $-\mathbb{I}$, which allows the removal of a control as $C^{(4)}(-\mathbb{I}) = C^{(3)}(-Z)$, with $-Z$ replacing one of the previous controls.
These types of matrix element-specific opportunities for circuit reduction will not be included in the upper-bound resource scaling calculations of Sec.~\ref{sec: Main: resource scaling}. 

\section{Resource scaling} \label{sec: Main: resource scaling}

Above, we have considered q-deformation for quantum simulation as a technique to improve circuit synthesis while maintaining proper convergence properties of field digitizations.
In order to assess the synthesis advantage, we evaluate a scalable upperbound to the circuit resources for a single q-deformed plaquette operator Trotter step. As suggested in Ref.~\cite{Zache:2023dko}, a baseline expression for the circuit resource cost $\xi_{\text{Trot}}$ of each plaquette-operator Trotter step is
\begin{equation}\label{eq: Trotter depth}
    \xi_{\text{Trot}} \leq 2(2\xi_{F_{1,2}} + \xi_{F_3} + \xi_G) + \xi_{\tilde{\Box}} \ \ \ ,
\end{equation}
where $\xi_{F_{1,2}}$ is the resource cost of the phased F-moves with four controls ($F_1$ and $F_2$), $\xi_{F_3}$ is the cost of $F_3$, $\xi_G$ is the cost of $G$, and $\xi_{\tilde{\Box}}$ is the cost of the 
diagonal plaquette time evolution operator acting and controlled on one qudit each.

We use a circuit model that assumes the availability of single-qudit Pauli unitaries ($X_{jk} = |j\rangle \langle k| + |k\rangle\langle j| + \sum_{\ell \neq j,k} |\ell\rangle\langle \ell|$, $Y_{jk} = i|k\rangle \langle j| - i |j\rangle \langle k| + \sum_{\ell \neq j,k} |\ell\rangle \langle \ell|$, and $Z_{jk} = |j\rangle\langle j| - |k\rangle\langle k| + \sum_{\ell \neq j,k} |\ell\rangle\langle \ell|$), two-level Givens rotations generated by exponentiating their non-unitary Hermitian counterparts ($\mathcal{X}_{jk} = |j\rangle \langle k| + |k\rangle\langle j|$, $\mathcal{Y}_{jk} = i|k\rangle \langle j| - i |j\rangle \langle k|$, and $\mathcal{Z}_{jk} = |j\rangle\langle j| - |k\rangle\langle k|$), and generalized controlled $X$ (GCX) gates \cite{Di:2013qvb}.  

\begin{figure} 
\label{fig: GCX resources}
    \centering
    \includegraphics[width=0.9\linewidth]{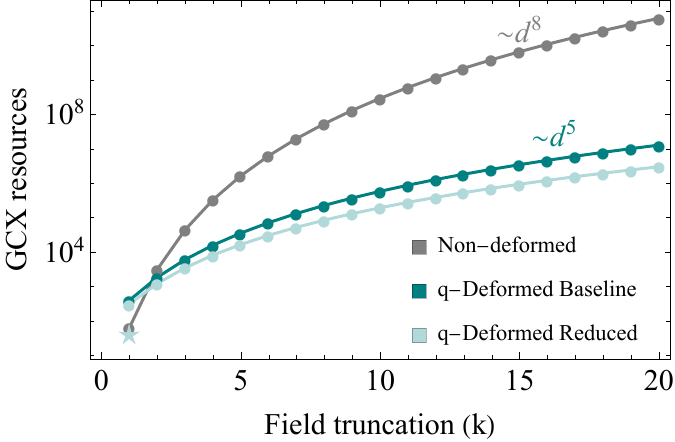}
    \caption{Explicit upperbounds to the number of GCX gates capable of performing one plaquette operator Trotter step as a function of local truncated Hilbert space dimension $d = k+1$ calculated via the expressions in Table~\ref{tab:gcxCalc} of Appendix~\ref{sec: Appendix: decomp}. For reference, the GCX scaling of the non-deformed theory as reported in Ref.~\cite{Jiang:2025ufg} is shown in gray. The baseline(reduced) GCX scaling of the q-deformed Trotter step is shown in dark teal(light teal). 
    For the qubit truncation $k = 1$, immediate improvement (light teal star) is available as discussed in Appendix~\ref{Appendix: circuits}.} 
\end{figure}
As one metric for quantifying circuit complexity, consider two-qudit entangling gates by specifying the general resource costs expression, Eq.~\eqref{eq: Trotter depth}, to GCX gates. 
Fig.~\ref{fig: GCX resources} shows the GCX scaling found when multi-controlled single-qudit gates are decomposed into single-qudit gates and GCX gates using an auxiliary qudit.
The dark teal line in Fig.~\ref{fig: GCX resources} is the result of the general scheme described in Appendix~\ref{sec: Appendix: decomp}, which uses a 5-dimensional auxiliary qudit. 
These basic techniques already accomplish $\mathcal{O}(d^5)$ GCX scaling, furnishing an improvement in the scaling of two-qudit entangling gates over both the $\mathcal{O}(d^8)$ non-deformed scaling found by Ref.~\cite{Jiang:2025ufg} (the gray line of Fig.~\ref{fig: GCX resources}, which uses a 4-dimensional auxiliary qudit) and the $\mathcal{O}(d^6)$ scaling estimated by Ref.~\cite{Zache:2023dko} with a 5-dimensional auxiliary qudit.

Beyond this general approach to decomposition of circuit elements, the light teal line in Fig.~\ref{fig: GCX resources} shows that it is possible to further reduce the GCX cost by taking advantage of GVC freedom and several structural features of the phased F-moves and diagonalizing G-move.
A primary contribution of the reduced scheme is its attention to the number of levels actively mixed by a single-qudit unitary. 
In general, the implementation resources of a single-qudit unitary acting on $m$ levels scales with $m$. Because F-moves only mix $m \leq \lceil\frac{d}{2}\rceil < d$ levels, it is economical to treat each single-qudit unitary as a unitary acting on the smaller $m$-dimensional subspace. 
That is, each circuit element present in Eq.~\eqref{eq: Trotter depth} is further broken down into a weighted sum over $m$-level components so that, for example, the cost of four-controlled phased F-moves is represented as $\xi_{F_{1,2}} = \sum_{m = 1}^{\lceil\frac{k+1}{2}\rceil}n_4(m,k)\xi_{C^{(4)}U_m}$. The weights $n(m,k)$ for each circuit component are determined in Appendix~\ref{Appendix: counting actively mixed} and the impact to circuitry is detailed in Appendix~\ref{sec: Appendix: decomp}. Additionally, in our reduced decomposition scheme, control sectors of $\Box'''$ are interleaved (as shown in Fig.~\ref{fig: gmove to gu}), eliminating the need for the diagonalizing $G$ unitaries to be controlled. The cost of this part of the circuit is thus represented together in terms of $\Box'''$ so that 
\begin{equation}
    \xi_{\text{Trot}} \leq 2(2\xi_{F_{1,2}} + \xi_{F_3}) + \xi_{\Box'''} \ \ \ .
\end{equation}
Finally, our reduced decomposition scheme also takes advantage of the antisymmetry of $\Box'''$ to halve the number of controlled rotations required to implement its time evolution.

The three orders of polynomial reduction in GCX resource scaling that q-deformation affords is intuitively consistent with the difference in Hilbert space extent between the 8-link plaquette operator and the 5-link (phased) F-move. 
That is, the largest operator to be decomposed in the non-deformed simulation is the plaquette operator with four control links and four active links, each with Hilbert space of dimension $d$. 
In the q-deformed simulation, the largest operator to be decomposed is a four-controlled (phased) F-move with four control links and one active link.

In both the baseline (dark teal) and reduced (light teal) approaches to circuitry, q-deformation offers an advantage in GCX resource scaling over the non-deformed simulation. While all quantum resource cost upperbounds are subject to reduction as quantum algorithms are discovered or as native gate sets evolve with hardware advances, it is interesting to note that significantly less optimization has been invested thus far in the q-deformed quantum compilation.

For $k=1$, the reduced q-deformed scheme uses 306 GCX gates for a q-deformed plaquette Trotter step, compared to the 62 GCX gates attributed to the non-deformed scheme of \cite{Jiang:2025ufg}. However, additional optimizations are available that have not been included in our explicit upperbound across $k$ truncations. 
For instance, using the same decomposition scheme but neglecting the presence of identity operators and reducing $\ell$-controlled phases to $(\ell-1)$-controlled unitaries as demonstrated in Fig.~\ref{fig: schematic circuit}, this cost can be reduced to 224 GCX gates. Alternatively, because each phased F-move in the $k=1$ case only requires rotations around a single axis ($X$ for the GVC in the right column of Table~\ref{tab:quditF1F2F3Gunitaries}), they can be implemented with uniformly controlled techniques \cite{Di:2013qvb,Jiang:2025ufg}, reducing the GCX cost to 80 GCX gates. 
By further modifying the GVC and computing intermediate parities, as demonstrated in Appendix~\ref{Appendix: circuits} and shown in Fig.~\ref{fig: reduced parity circuit}, these estimates can be reduced to 48 GCX gates (light teal star in Fig.~\ref{fig: GCX resources}).
Thus, q-deformation is found to reduce current GCX cost for all values of the $k$ truncation.

\section{Discussion}\label{sec: Main: discussion}

Q-deforming the gauge group of an LGT simulation generates an infinite tower of finite groups that has been observed to maintain a smooth trajectory to the field continuum.  
For example, though the gauge invariance condition is strengthened through the deformation to eliminate configurations of high flux density, the physical subspace of the plaquette operator continues to grow comparably to that of the non-deformed theory under the same truncation.

In this work, we have constructed a path for simulating the q-deformed SU(2) Yang Mills lattice gauge theory at arbitrary truncations, with techniques that are expected to extend to SU(3)~\cite{Ciavarella:2021nmj,Hayata:2023bgh}. 
Building on Ref.~\cite{Zache:2023dko}, we have computed phased F-moves that diagonalize this plaquette operator and presented a diagrammatic approach to identifying F-sequence circuit components.
We present a systematic choice of GVC that streamlines the construction of valid quantum gates that are unitary over the full computational Hilbert space. 
While we have focused on a 1+1D plaquette chain, these techniques are naturally extendable to 2+1D and 3+1D systems via additional steps in the F-sequence~\cite{Zache:2023dko}.

Employing these techniques, we have provided an explicit upperbound calculation reducing the GCX resource scaling of a q-deformed plaquette Trotter step by $\mathcal{O}(d^3)$ over current optimizations of the non-deformed simulation~\cite{Jiang:2025ufg}. 
This improvement is enabled by the F-sequence diagonalization and the first choice of GVC, reducing the fundamental operator size from 8 to 5 links.
Using both a baseline circuit decomposition scheme and a reduced scheme that leverages additional structure found in each circuit component, we find a GCX resource advantage for all truncations.
As seen for $d=2$, further reductions to the q-deformed resource requirements are anticipated to be available---perhaps scaling polynomially with the number of qudits---by considering more advanced circuit synthesis techniques~\cite{Bacon:2004moa,Di:2013qvb,Gokhale:2019ujm,Litteken:2022uip,Murairi:2024xpc,Jiang:2025ufg}. 

We have focused on GCX resources as one approach to quantifying circuit complexity. However, different quantum architectures may yield dominant costs for alternate elements of quantum compilation. 
For example, fault-tolerant devices can require expensive protocols to implement highly controlled operators and non-Clifford gates over logical registers.
In these contexts, T-gates (rather than GCX gates) are a typical focus of resource evaluation~\cite{Fowler:2012wtt,Kan:2021xfc,Gustafson:2024kym,Rhodes:2024zbr,Ciavarella:2025bsg,Perez:2025cxl,Balaji:2025afl}. However, practical cost models may evolve as fault-tolerant architectures develop~\cite{Litinski:2019uvg, Kim:2025guu}. 
By providing a complete unitary implementation strategy of time evolution operators for q-deformed SU(2) lattice pure-gauge theory, the present work establishes a concrete foundation to support optimizations for architecture-specific circuit synthesis.

Q-deformed LGT simulations may be of further interest for error correction as finite (sub)group approaches may facilitate 
greater robustness of gauge invariance to quantum noise~\cite{Gustafson:2023swx}, an important consideration for quantum simulations~\cite{Stryker:2018efp, Lamm:2020jwv, Rajput:2021trn, Mathew:2022nep, Halimeh:2022mct, Spagnoli:2024mib, Spagnoli:2026qni}.
Furthermore, the finite group structure, enabled at all truncations by q-deformation, yields tools for systematic analysis, e.g., of noise sensitivities, physical dimensionalities, and circuit resources.
The presented concrete foundation also supports future work elaborating these connections.

\begin{acknowledgments}
The authors thank Olivia Di Matteo for discussions of circuit synthesis and thank Torsten Zache for discussions at early stages of this work facilitated by the Munich Conference on Quantum Science and Technology (MCQST) funded by the Deutsche Forschungsgemeinschaft (DFG, German Research Foundation) under Germany's Excellence Strategy - EXC2111 - 390814868. 
ZWM acknowledges support from the National Science Foundation (NSF) Graduate Research Fellowship Program (GRFP) under grant number DGE\nobreakdash-2039655.  NK acknowledges support from the
NSF STAQ Program (PHY-2325080).
\end{acknowledgments}

\bibliography{biblio}

\newpage
\onecolumngrid 
\appendix

\section{Definition and properties of F-symbols}\label{sec: Appendix: F def and props}
\subsection{Definition}\label{sec: Appendix: F definition}
Q-deformed F-symbols as defined in Eq.~\eqref{eq: q-deformed F matrix} can be calculated by deforming the Racah formula for Wigner $6j$ symbols,
\begin{equation} \label{eq: Racah 6j}
\begin{gathered}
    \begin{Bmatrix} 
      a & b & e \\ 
      c & d & f
    \end{Bmatrix}_k = \sqrt{\Delta_{abe}\Delta_{adf}\Delta_{cbf}\Delta_{cde}} \sum_{J=J_{\text{min}}}^{J_{\text{max}}}(-1)^{J}[J+1]!
    \\
    \times \left(\prod_{n = 1}^{3}[T_n - J]!\right)^{-1}\left(\prod_{m = 1}^{4}[J - \tau_{m}]!\right)^{-1} \ \ \ ,
\end{gathered}
\end{equation}
\noindent where the q-deformation parameter $k$ subscript on the right side have been suppressed as in the main text, $T_{n}$ are the three tetrads
$ T = \{ a+b+c+d, a+c+e+f, b+d+e+f\}$, and $\tau_{m}$ are the four triads
$ \tau = \{a+b+e, a+d+f, c+b+f, c+d+e\}$. The bounds of the sum, $J_{\text{min}}$ and $J_{\text{max}}$, are:
\begin{subequations}
    \begin{align}
        J_{\text{min}} &= \text{max} \left(\tau\right) \\
        J_{\text{max}} &= \text{min}\left( T\right)  \ \ \ .
    \end{align}
\end{subequations}
For the F-moves, each of the triads is associated with a three-point vertex on either the original or modified lattice. The q-deformed triangle deltas are associated with the triads as 
\begin{equation}\label{eq: triangle delta def}
    \Delta_{abc} := \delta_{abc} \frac{[a+b-c]![a-b+c]![-a+b+c]!}{[a+b+c+1]!} \ \ \ ,
\end{equation}
where $\delta_{abc}$ is an admissibility function such that $\delta_{abc} = 1$ if flux at vertex $a,b,c$ satisfies the q-deformed gauge singlet conditions of Eqns.~\eqref{eq: gauge singlet non-def} and~\eqref{eq: fusion constraint}, and $\delta_{abc} = 0$ otherwise. Each of the factorials in the above formulae are promoted to q-deformed factorials of Eq.~\eqref{eq: q-number} q-numbers,
\begin{equation}
    [n]! = 
    \begin{cases}
    1 & \text{if } n=0 \\
    \prod_{m = 1}^{n} [m] & \text{if } n \neq 0
    \end{cases} \ \ \ .
\end{equation}
In total, Eq.~\eqref{eq: Racah 6j} is equivalent to the standard $6j$ definition but with all scalars q-deformed (indicated by square brackets) and with the additional fusion constraint Eq.~\eqref{eq: fusion constraint} included in the admissibility function.

To see one connection between this choice of q-deformation and the fusion constraint, consider some triad $a,b,c$ that satisfies the usual gauge singlet constraints but not the q-deformed fusion constraint. For $a,b,c$ satisfying $a+b+c \in \mathbb{N}_0$, then $a + b + c = k + 1$ is the minimal violation of the fusion rule. Under the definition of quantum numbers in Eq.~\eqref{eq: q-number}, the factor of $[a+b+c+1]$ that appears in the factorial in the denominator of Eq.~\eqref{eq: triangle delta def} vanishes at this minimal violation,
\begin{equation}
[k+2] = \frac{\text{sin}\left( \frac{\pi}{k+2}(k+2)\right)}{\text{sin}\left(\frac{\pi}{k+2} \right)} = 0  \ \ \ . \nonumber
\end{equation}
By extension, any $[a+b+c+1]!$ where $a+b+c > k$ is $0$. Therefore, without the q-deformed fusion constraint within the admissibility function, triads violating this fusion constraint produce an undefined triangle coefficient.

\subsection{Properties of F-symbols}\label{sec: Appendix: properties}
The following section summarizes relevant properties of F-symbols~\cite{Biedenharn1995, Zache:2023dko, Kirillov:1991ec}. From the tetrahedral symmetry of the $6j$ symbol, q-deformed F-symbols inherit several symmetries. F-symbols are invariant under exchange of the first two columns,
\begin{equation}
 \begin{bmatrix}
        a&b&e\\
        c&d&f
    \end{bmatrix}
    =
    \begin{bmatrix}
        b&a&e\\
        d&c&f
    \end{bmatrix}
    \ \ \ ,
\end{equation}
and inverting any two columns at a time, 
\begin{equation}
 \begin{bmatrix}
        a&b&e\\
        c&d&f
    \end{bmatrix}
    =
    \begin{bmatrix}
        c&d&e\\
        a&b&f
    \end{bmatrix}
    =
    \begin{bmatrix}
        c&b&f\\
        a&d&e
    \end{bmatrix}
    =
    \begin{bmatrix}
        a&d&f\\
        c&b&e
    \end{bmatrix}
    \ \ \ .
\end{equation}
Exchanging the final column with the first or second column introduces a dimension-dependent scale factor, 
\begin{equation}
\begin{bmatrix}
        a&b&e\\
        c&d&f
    \end{bmatrix}
    =
   \frac{v_e v_f}{v_a v_c} \begin{bmatrix}
        e&b&a\\
        f&d&c
    \end{bmatrix}\ \ \ ,
\end{equation}
with $v_j = (-1)^{-j} \sqrt{D(j)}$, though this can be avoided in the present calculations. Additionally, F-symbols obey the following orthogonality relation,
\begin{equation} \label{eq: F orthogonality relation}
\sum_{J}
\begin{bmatrix} 
      j_{1} & j_{2} & J \\ 
      j_{3} & j_{4} & j'
\end{bmatrix}
\begin{bmatrix} 
      j_{1} & j_{2} & J \\ 
      j_{3} & j_{4} & j
\end{bmatrix} 
= \delta_{jj'} \ \ \ ,
\end{equation}
and the pentagon identity,
\begin{equation} \label{eq: F pentagon identity}
\sum_{J}
    \begin{bmatrix} 
      j_{1} & j_{2} & j_{5} \\ 
      j_{3} & j_{4} & J
    \end{bmatrix}
    \begin{bmatrix} 
      j_{6} & j_{7} & j_{4} \\ 
      J & j_{1} & j_{8}
    \end{bmatrix} 
    \begin{bmatrix} 
      j_{8} & j_{7} & J \\ 
      j_{3} & j_{2} & j_{9}
    \end{bmatrix} 
= 
\begin{bmatrix} 
      j_{1} & j_{2} & j_{5} \\ 
      j_{9} & j_{6} & j_{8}
    \end{bmatrix}
    \begin{bmatrix} 
      j_{6} & j_{7} & j_{4} \\ 
      j_{3} & j_{5} & j_{9}
    \end{bmatrix} \ \ \ . 
\end{equation}
Because F-symbols are real, 
$\begin{bmatrix}
        a&b&e\\
        c&d&f
    \end{bmatrix}^\dagger
 = \begin{bmatrix}
        a&b&f\\
        c&d&e
    \end{bmatrix}$, 
the unitarity of q-deformed F-moves in the q-deformed physical subspace that satisfies the fusion constraint follows directly from the choice of deformation parameter (Eq.~\eqref{eq: deformation parameter}) and resulting orthogonality relation. Outside this q-deformed physical subspace, the  admissibility function causes the F-moves to annihilate all states.
Extending unitarity across the computational Hilbert space is one function of the GVCs discussed in Section~\ref{sec: Main: circuitry}.

\section{Diagonalizaton of the plaquette operator}\label{sec: Appendix: diagonalization}

\subsection{Phased F-move sequence}
In the following, the diagonalization process is continued for $F_2$ and $F_3$ as presented for $F_1$ in the main text, applying the pentagon identity and orthogonality relation at each step. 

By inspection of Fig.~\ref{fig: f sequence}, $F_2$ is expected to have the form
\begin{equation}\label{eq: F2 ansatz}
    \langle ...Q_r...| F_2| ...q_r...\rangle = (-1)^{\alpha_2}
    \begin{bmatrix}
        j_a^t & j_a^b & Q_r\\
        j_r^b & j_r^t & q_r
    \end{bmatrix} \ \ \ .
\end{equation}
The plaquette operator after the action of $F_2$ is
\begin{multline}
\label{eq: F2 transformation abbrev.}
    \langle ...j_a^{t \prime} j_a^{b \prime} Q_r'...|F_2 F_1\Box F_1^\dagger F_2^{\dagger}|...j_a^t j_a^b Q_r...\rangle 
    \\
    = 
    \textcolor{Purple!90!Black}{\delta_{j_l^t j_l^b Q_l}} \sum_{q_r, q_r'} (-1)^{-\Delta j_a^t - \Delta j_a^b + \Delta q_r + \alpha_2' - \alpha_2}
    \begin{bmatrix} 
          j_r^t & j_r^b & Q_r' \\
          j_a^{b \prime} & j_a^{t \prime} & q_r'
    \end{bmatrix}
    \begin{bmatrix} 
        1/2 & j_a^{b \prime} & j_a^b \\ 
        Q_l & j_a^t & j_a^{t \prime}
    \end{bmatrix}
    \\
    \times
    \begin{bmatrix} 
          j_{r}^{t} & q_{r} & j_{a}^{t} \\ 
          1/2 & j_{a}^{t\prime} & q_r'
    \end{bmatrix}
    \begin{bmatrix} 
          j_{r}^{b} & j_{a}^{b} & q_r \\ 
          1/2 & q_r' & j_{a}^{b\prime}
    \end{bmatrix} 
    \begin{bmatrix} 
        j_r^t & j_r^b & Q_r \\ 
        j_a^b & j_a^t & q_r
    \end{bmatrix}   \ \ \ ,
\end{multline}
where, as in the main text, capital indices indicate links on the modified lattice. As in Section~\ref{sec: qdefDiagF1}, an additional admissibility function, which will later be neglected under our choice of GVC, appears on the RHS. Throughout this section, we will highlight admissibility functions removed by this GVC in purple. By setting $\alpha_2' = -Q_r'-q_r'$ and $\alpha_2 = -q_r - Q_r$, the phase in the sum becomes $-\Delta j_a^t - \Delta j_a^b - \Delta Q_r$, which has no dependence on the summed indices, allowing application of Eq.~\eqref{eq: F pentagon identity} and Eq.~\eqref{eq: F orthogonality relation}. The sum over $q_r'$ may be addressed by identifying $J = q_r'$ and $\mathbf{j} = \{ 1/2, j_a^{t \prime}, j_r^t, q_r, j_a^t, j_a^b, j_r^b, j_a^{b \prime}, Q_r' \}$ in the pentagon identity of Eq.~\eqref{eq: F pentagon identity}, leading to   
\begin{multline}
    \langle ...j_a^{t \prime} j_a^{b \prime} Q_r'...|F_2 F_1\Box F_1^\dagger F_2^{\dagger}|...j_a^t j_a^b Q_r...\rangle 
    \\
    = 
    \textcolor{Purple!90!Black}{\delta_{j_l^t j_l^b Q_l}} \sum_{q_r} (-1)^{-\Delta j_a^t - \Delta j_a^b - \Delta Q_r}
    \begin{bmatrix} 
        1/2 & j_a^{b \prime} & j_a^b \\ 
        Q_l & j_a^t & j_a^{t \prime}
    \end{bmatrix}
    \begin{bmatrix} 
      1/2 & j_a^{t \prime} & j_a^t \\
      Q_r' & j_a^b & j_a^{b \prime}
    \end{bmatrix}
    \begin{bmatrix} 
        j_a^b & j_r^b & q_r \\
        j_r^t & j_a^t & Q_r'
    \end{bmatrix}
    \begin{bmatrix} 
        j_r^t & j_r^b & Q_r \\ 
        j_a^b & j_a^t & q_r
    \end{bmatrix}   \ \ \ . 
\end{multline}
Finally, the sum over $q_r$ may be performed to produce $\delta_{Q_r Q_r'}$ via the orthogonality relation, resulting in 
\begin{equation}
    \langle ...j_a^{t \prime} j_a^{b \prime}...|F_2 F_1\Box F_1^\dagger F_2^{\dagger}|...j_a^t j_a^b...\rangle 
    \\
    = \textcolor{Purple!90!Black}{\delta_{j_l^t j_l^b Q_l} \delta_{j_r^t j_r^b Q_r}}(-1)^{-\Delta j_a^t - \Delta j_a^b - \Delta Q_r}
    \begin{bmatrix} 
        1/2 & j_a^{b \prime} & j_a^b \\ 
        Q_l & j_a^t & j_a^{t \prime}
    \end{bmatrix}
    \begin{bmatrix} 
      1/2 & j_a^{t \prime} & j_a^t \\
      Q_r & j_a^b & j_a^{b \prime}
    \end{bmatrix}   \ \ \ ,
\end{equation}
where again we have inserted an admissibility function into the RHS in order for the equality to hold over the full computational Hilbert space containing both unphysical and physical basis states. As discussed in Section~\ref{sec: qdefDiagF1}, the accompanying stability granted to each register in these admissibility functions allows them to be safely removed by our GVC, causing the plaquette operator to span two fewer qudits with each step of the diagonalization.

Similarly, the ansatz for $F_3$ follows from Fig.~\ref{fig: f sequence} as
\begin{equation}\label{eq: F3 ansatz}
    \langle ...J_a^{t}...| F_3| ...j_a^t...\rangle = (-1)^{\alpha_3}
    \begin{bmatrix}
        q_l & q_r & J_a^{t}\\
        j_a^b & j_a^b & j_a^t
    \end{bmatrix} \ \ \ .
\end{equation}
After the action of $F_3$, the plaquette operator matrix elements are
\begin{multline}
\label{eq: F3 transformation}
    \langle ...J_a^{t \prime} j_a^{b \prime}...|F_3F_2F_1\Box F_1^\dagger F_2^\dagger F_3^\dagger|...J_a^t j_a^b...\rangle 
    \\
    =
    \textcolor{Purple!90!Black}{\delta_{j_l^t j_l^b Q_l} \delta_{j_r^t j_r^b Q_r}}
    \sum_{j_a^t j_a^{t \prime}}
    (-1)^{-\Delta j_a^t - \Delta j_a^b + \alpha_3' - \alpha_3}
    \begin{bmatrix}
        Q_l & Q_r & J_a^{t \prime}\\
        j_a^{b \prime} & j_a^{b \prime} & j_a^{t \prime}
    \end{bmatrix}
    \begin{bmatrix} 
        1/2 & j_a^{b \prime} & j_a^b \\ 
        Q_l & j_a^t & j_a^{t \prime}
    \end{bmatrix}
    \\
    \times
    \begin{bmatrix} 
      1/2 & j_a^{t \prime} & j_a^t \\
      Q_r & j_a^b & j_a^{b \prime}
    \end{bmatrix} 
    \begin{bmatrix}
        Q_l & Q_r & j_a^t\\
        j_a^b & j_a^b & J_a^t
    \end{bmatrix} \ \ \ .
\end{multline}
Letting $\alpha_3' = J_a^{t \prime} + j_a^{t \prime}$ and $\alpha_3 = j_a^t + J_a^t$, the exponent becomes $\Delta J_a^t - \Delta j_a^b$ so that it is no longer dependent on the summed indices. As before, by identifying $J = j_a^{t \prime}$ and $\mathbf{j} = \{ Q_r, Q_l, j_a^{b \prime}, j_a^{b \prime}, J_a^{t \prime}, j_a^b, 1/2, j_a^t, j_a^b\}$, the pentagon identity performs the sum over $j_a^{t \prime}$, leading to
\begin{multline}
    \langle ...J_a^{t \prime} j_a^{b \prime}...|F_3F_2F_1\Box F_1^\dagger F_2^\dagger F_3^\dagger|...J_a^t j_a^b...\rangle  
    \\
    =
    \textcolor{Purple!90!Black}{\delta_{j_l^t j_l^b Q_l} \delta_{j_r^t j_r^b Q_r}}
    \sum_{j_a^t}
    (-1)^{\Delta J_a^t - \Delta j_a^b}
    \begin{bmatrix} 
          Q_r & Q_l & J_a^{t \prime}\\
          j_a^b & j_a^b & j_a^{t}
    \end{bmatrix}
    \begin{bmatrix} 
          j_a^b & 1/2 & j_a^{b \prime} \\
          j_a^{b \prime} & J_a^{t \prime} & j_a^b
    \end{bmatrix}
    \begin{bmatrix}
        Q_l & Q_r & j_a^t\\
        j_a^b & j_a^b & J_a^t
    \end{bmatrix} \ \ \ .
\end{multline}
Finally, the orthogonality relation produces a $\delta_{J_a^t J_a^{t'}}$ upon performance of the sum over $j_a^t$,
\begin{equation}
    \langle ... j_a^{b \prime}...|F_3F_2F_1\Box F_1^\dagger F_2^\dagger F_3^\dagger|... j_a^b...\rangle 
    \\
    =
     \textcolor{Purple!90!Black}{\delta_{j_l^t j_l^b Q_l} \delta_{j_r^t j_r^b Q_r} \delta_{Q_l Q_r J_a^t}}
    (-1)^{-\Delta j_a^b}
    \begin{bmatrix} 
          j_a^b & 1/2 & j_a^{b \prime} \\
          j_a^{b \prime} & J_a^{t} & j_a^b
    \end{bmatrix} \ \ \ ,
\end{equation}
where again the relevant admissibility function has been introduced to the RHS so the expression applies throughout the computational Hilbert space.
Because these physicality criteria have been fixed by corresponding $\delta_{jj'}$ factors, we may choose a non-trivial GVC in the unphysical space that removes them, resulting in Eq.~\eqref{eq: plaq after F3},
\begin{equation}
    \langle ...j_a^{b \prime}...|\Box'''|...j_a^b...\rangle 
    \\
    =
    (-1)^{-\Delta j_a^b}
    \begin{bmatrix} 
          j_a^b & 1/2 & j_a^{b \prime} \\
          j_a^{b \prime} & J_a^{t} & j_a^b
    \end{bmatrix} \ \ \ ,
\end{equation}
with $\Box''' = F_3F_2F_1\Box F_1^\dagger F_2^\dagger F_3^\dagger$. Here, the plaquette operator has been reduced to a single-qudit operator on the $|j_a^b\rangle$ register controlled only on the transformed $|j_a^t\rangle$ register.

Notice that, relative to the diagonalization described in \cite{Zache:2023dko}, the F-moves that diagonalize the plaquette operator for the SU(2) Hamiltonian with plaquette operator of Eq.~\eqref{eq: plaquette F-symbols} carry a phase in addition to the F-matrix element. While F-symbols themselves are real, phased F-moves have non-real elements.

\subsection{Flux hierarchy inversion symmetry of \texorpdfstring{$\Box'''$}{B'''}}\label{Appendix: FHI symmetry}
Due to the structure of the F-symbol in 
Eq.~\eqref{eq: plaq after F3},
the transformed plaquette operator $\Box'''$ has an additional \textit{flux hierarchy inversion} symmetry, meaning that $\langle j_a^{b\prime}|\Box'''|j_a^b\rangle$ is invariant under $j_a^b \rightarrow k/2 - j_a^{b\prime}$ and $j_a^{b \prime} \rightarrow k/2 - j_a^{b}$, which exchanges high and low flux.  For example, with $J_a^t = 1$ and $k = 7$, $\langle \frac{1}{2}|\Box'''|1\rangle = \langle \frac{5}{2}|\Box'''|3\rangle $. This corresponds to persymmetry (symmetry about the antidiagonal) of the matrix representation of $\Box'''$ in the $|j_a^b\rangle$ basis. The triangle deltas associated with the triads in $\Box'''$ give the constraint $|j_a^{b \prime} - j_a^b| = 1/2$. Using this constraint and letting $j_a^{b \prime} = j_a^b + 1/2$ and $j_a^b < \frac{\lfloor \frac{k}{2}\rfloor}{2}$ (i.e., selecting a low-flux pair) for the sake of simplicity, the inversion transformation can be rendered as
\begin{equation}
    \begin{gathered}
    j_a^b \rightarrow j_a^b + \gamma \\
    j_a^{b \prime} \rightarrow j_a^{b \prime} + \gamma
    \end{gathered} \ \ \ ,
\end{equation}
where $\gamma = \frac{k-1}{2} - 2j_a^b$. Under this transformation, the F-symbol in $\Box'''$ becomes 
\begin{equation}
\label{eq:fluxhierarchyinversionF}
    \begin{bmatrix}
        j_a^b & 1/2 & j_a^{b \prime} \\
        j_a^{b \prime} & J_a^t & j_a^b
    \end{bmatrix} 
    \rightarrow
    \begin{bmatrix}
        j_a^b + \gamma & 1/2 & j_a^{b \prime} + \gamma \\
        j_a^{b \prime} + \gamma & J_a^t & j_a^b + \gamma
    \end{bmatrix} \ \ \ .
\end{equation}
Recall that an F-symbol is defined by three parts as in Eq.~\eqref{eq: q-deformed F matrix}: a phase, a dimension factor, and a $6j$ symbol. Here, we work part-by-part to prove that $\Box'''$ is invariant under the flux hierarchy inversion transformation.

First, consider the $6j$ part. As defined in Eq.~\eqref{eq: Racah 6j}, the $6j$ symbol has two parts: a triangle delta term and a sum over $J$ that includes both tetrad and triad terms. We will first consider the triangle delta term $\sqrt{\Delta_{abe} \Delta_{adf} \Delta_{cbf} \Delta_{cde}}$. With $\Delta_{abc}$ defined as in Eq.~\eqref{eq: triangle delta def}, this term transforms as
\begin{equation}
    \Delta_{j_a^b \frac{1}{2}j_a^{b \prime}}^2\Delta_{j_a^b J_a^t j_a^b} \Delta_{j_a^{b \prime} J_a^t j_a^{b \prime}} \rightarrow \Delta_{(j_a^b+\gamma) \frac{1}{2} (j_a^{b \prime}+\gamma)}^2\Delta_{(j_a^b+ \gamma)J_a^t (j_a^b + \gamma)} \Delta_{(j_a^{b \prime} + \gamma) J_a^t (j_a^{b \prime} + \gamma)} \ \ \ .
    \label{eq:transform4Delta}
\end{equation}
Under the transformation, the first delta $\Delta_{j_a^b \frac{1}{2} j_a^{b \prime}}$ becomes
\begin{align}
    &\Delta_{(j_a^b + \gamma) \frac{1}{2} (j_a^{b \prime}+\gamma)} \nonumber \\
     &\qquad= \delta_{(j_a^b + \gamma) \frac{1}{2} (j_a^{b \prime}+\gamma)} \frac{[j_a^b + \gamma + 1/2 - (j_a^{b \prime} + \gamma)]![j_a^b + \gamma - 1/2 + j_a^{b \prime} + \gamma]![-(j_a^b + \gamma)+1/2+j_a^{b \prime} + \gamma]!}{[j_a^b + \gamma + 1/2 + j_a^{b \prime} + \gamma + 1]!} \\
    &\qquad= \delta_{(j_a^b + \gamma) \frac{1}{2} (j_a^{b \prime}+\gamma)} \frac{[0]![2j_a^b + 2\gamma]![1]!}{[2j_a^b + 2\gamma + 2]!} \\
    &\qquad= \delta_{(j_a^b + \gamma) \frac{1}{2} (j_a^{b \prime}+\gamma)} \frac{[k-2j_a^b - 1]![1]}{[k-2j_a^b + 1]!} \\
    &\qquad= \delta_{(j_a^b + \gamma) \frac{1}{2} (j_a^{b \prime}+\gamma)} \frac{[1]}{[k-2j_a^b][k-2j_a^b + 1]} \ \ \ .
\end{align}
Note that following Eq.~\eqref{eq: q-number} the q-numbers themselves have an inversion symmetry,
\begin{equation}\label{eq: q-number under flux inversion}
    [n]_k = [k+2 - n]_k  \ \ \ ,
\end{equation}
because $\text{sin}(\frac{\pi n}{k+2}) = \text{sin}(\pi - \frac{\pi n}{k+2})$. Using this fact, $\Delta_{(j_a^b + \gamma) \frac{1}{2} (j_a^{b \prime}+\gamma)}$ can be rewritten as
\begin{equation}
    \Delta_{(j_a^b + \gamma) \frac{1}{2} (j_a^{b \prime}+\gamma)} = \delta_{(j_a^b + \gamma) \frac{1}{2} (j_a^{b \prime}+\gamma)} \frac{[1]}{[2j_a^b + 2][2j_a^b + 1]} \ \ \ .
\end{equation}
Now, notice that the non-transformed triangle delta $\Delta_{j_a^b \frac{1}{2} j_a^{b \prime}}$ contains
\begin{equation}
    \frac{[j_a^b + 1/2 - j_a^{b \prime}]![j_a^b - 1/2 + j_a^{b \prime}]![-j_a^b + 1/2 + j_a^{b \prime}]!}{[j_a^b + 1/2 + j_a^{b \prime} + 1]!} = \frac{[0]![2j_a^b]![1]!}{[2j_a^b + 2]!} = \frac{[1]}{[2j_a^b + 1][2j_a^b + 2]} \ \ \ ,
\end{equation}
so the numerical factors agree. Now, consider the admissibility function $\delta_{(j_a^b + \gamma) \frac{1}{2}(j_a^{b \prime} + \gamma)}$ that enforces the gauge singlet conditions and fusion constraint of Eqns.~\eqref{eq: gauge singlet non-def} and~\eqref{eq: fusion constraint}. Upon transformation, those constraints are preserved in aggregate
\begin{equation}
\begin{alignedat}{2}
\tikzmark{third}j_a^b + j_a^{b \prime} \geq 1/2 \quad &\rightarrow\quad  j_a^b + j_a^{b \prime} + 2\gamma \geq 1/2 
    \quad &&\iff\quad j_a^b+j_a^{b \prime}+1/2\leq k \\
     j_a^b + 1/2 \geq j_a^{b \prime}  \quad &\rightarrow\quad j_a^b + \gamma + 1/2 \geq j_a^{b \prime} + \gamma 
     \quad &&\iff\quad j_a^b + 1/2 \geq j_a^{b \prime} \tikzmark{first}   \\
     \tikzmark{fourth}j_a^{b \prime}  + 1/2 \geq j_a^b  \quad &\rightarrow\quad j_a^{b \prime} + \gamma + 1/2 \geq j_a^b + \gamma 
     \quad&&\iff\quad j_a^{b \prime} + 1/2 \geq j_a^b\\
     j_a^b + j_a^{b \prime} + 1/2 \leq k \quad &\rightarrow\quad j_a^b + j_a^{b \prime} + 2\gamma + 1/2 \leq k 
    \quad&&\iff\quad j_a^b + j_a^{b \prime} \geq 1/2  \tikzmark{second}\\
     j_a^b + j_a^{b \prime} + 1/2 \in \mathbb{N}_0 \quad &\rightarrow\quad j_a^b + j_a^{b \prime} + 2\gamma + 1/2 \in \mathbb{N}_0 
     \quad&&\iff\quad j_a^b + j_a^{b \prime} + 1/2 \in \mathbb{N}_0  \ \ \ ,
\end{alignedat}
\end{equation}
\begin{tikzpicture}[overlay, remember picture]
\draw [decoration={brace,amplitude=0.4em},decorate,very thick,black]
(first.north east) --  (second.east);
\draw [decoration={brace,amplitude=0.4em,mirror},decorate,very thick,black]
(third.north west) --  (fourth.west);
\end{tikzpicture}
i.e., $\delta_{j_a^b \frac{1}{2} j_a^{b \prime}} = \delta_{(j_a^b + \gamma) \frac{1}{2} (j_a^{b \prime} + \gamma)}$. The triangle inequalities before and after are bracketed for illustration.  Thus, the first triangle delta of Eq.~\eqref{eq:transform4Delta} is invariant under the flux inversion transformation, $\Delta_{(j_a^b + \gamma)\frac{1}{2}(j_a^{b \prime}+\gamma)} = \Delta_{j_a^b \frac{1}{2} j_a^{b \prime}}$. The next triangle delta transforms non-trivially,
\begin{equation}
\begin{aligned}
    \Delta_{(j_a^b + \gamma)J_a^t (j_a^b + \gamma)}
    &= \delta_{(j_a^b + \gamma)J_a^t (j_a^b + \gamma)}\frac{[2j_a^b + 2\gamma - J_a^t]![J_a^t]![J_a^t]!}{[2j_a^b + 2\gamma + J_a^t + 1]!} \\
    &=\delta_{(j_a^b + \gamma)J_a^t (j_a^b + \gamma)} \frac{[k-1-2j_a^b-J_a^t]![J_a^t]!^2}{[k-2j_a^b + J_a^t]!} \\
    &=\delta_{(j_a^b + \gamma)J_a^t (j_a^b + \gamma)} \frac{[J_a^t]!^2}{[k-2j_a^b - J_a^t]\ldots[k-2j_a^b + J_a^t]} \\
    &=\delta_{(j_a^b + \gamma)J_a^t (j_a^b + \gamma)} \frac{[J_a^t]!^2}{[2j_a^b + J_a^t + 2]\ldots[2j_a^b - J_a^t + 2]} \\
    &= \delta_{(j_a^b + \gamma)J_a^t (j_a^b + \gamma)} \frac{[J_a^t]!^2}{[2j_a^{b \prime} + J_a^t + 1]\ldots[2j_a^{b \prime} - J_a^t + 1]} \\
    &= \delta_{(j_a^b + \gamma)J_a^t (j_a^b + \gamma)}\frac{[J_a^t]!^2 [2j_a^{b \prime} - J_a^t]!}{[2j_a^{b \prime} + J_a^t + 1]!} \ \ \ ,
\end{aligned}
\end{equation}
where Eq.~\eqref{eq: q-number under flux inversion} has been used to rewrite the q-numbers containing $k$. The transformation of the admissibility function $\delta_{j_a^bJ_a^tj_a^b} \rightarrow \delta_{(j_a^b + \gamma)J_a^t(j_a^b + \gamma)}$ manifests in the singlet and fusion constraints as
\begin{equation}
\begin{alignedat}{2}
     \tikzmark{upperleft}2j_a^b \geq J_a^t\quad&\rightarrow\quad 2j_a^b + 2\gamma \geq J_a^t \quad&&\iff\quad 2j_a^{b \prime} + J_a^t \leq k \\
     \tikzmark{lowerleft}\phantom{j_a^b}J_a^t \geq 0 \quad&\rightarrow \quad  J_a^t \geq 0 &&\phantom{\iff\quad 2j_a^{b \prime} \geq J_a^t \ }\tikzmark{upperright}\\
     2j_a^b  + J_a^t \leq k \quad&\rightarrow\quad 2j_a^b + 2\gamma + J_a^t \leq k \quad&&\iff\quad 2j_a^{b \prime} \geq J_a^t \tikzmark{lowerright}\\
     J_a^t \in \mathbb{N}_0\quad&\rightarrow\quad J_a^t \in \mathbb{N}_0 \ \ \ ,
\end{alignedat}
\end{equation}
\begin{tikzpicture}[overlay, remember picture]
\draw [decoration={brace,amplitude=0.3em},decorate,very thick,black]
(upperright.north east) --  (lowerright.east);
\draw [decoration={brace,amplitude=0.3em,mirror},decorate,very thick,black]
(upperleft.north west) --  (lowerleft.west);
\end{tikzpicture}
which are the constraints imposed by $\delta_{j_a^{b \prime} J_a^t j_a^{b \prime}}$, once again with the fusion constraint benignly exchanged with one of the triangle inequalities. Thus,
\begin{equation}
    \Delta_{(j_a^b + \gamma) J_a^t (j_a^b + \gamma)} = \delta_{j_a^{b\prime} J_a^t j_a^{b \prime}}\frac{[J_a^t]!^2 [2j_a^{b \prime} - J_a^t]!}{[2j_a^{b \prime} + J_a^t + 1]!} = \Delta_{j_a^{b \prime} J_a^t j_a^{b \prime}}  \ \ \ .
\end{equation}
The transformation on this triangle delta has resulted in $j_a^b \rightarrow j_a^{b \prime}$. By similar logic, it can be shown that the transformation on the remaining triangle delta is captured by $j_a^{b \prime} \rightarrow j_a^b$, such that $\Delta_{(j_a^{b \prime} + \gamma) J_a^t (j_a^{b \prime} + \gamma)} = \Delta_{j_a^b J_a^t j_a^b}$. Together, these calculations identify the transformed expression at the right of Eq.~\eqref{eq:transform4Delta} with the original product on the left,
\begin{equation}
    \Delta_{(j_a^b+\gamma) \frac{1}{2} (j_a^{b \prime}+\gamma)}^2
    \Delta_{(j_a^{b \prime} + \gamma) J_a^t (j_a^{b \prime} + \gamma)}
    \Delta_{(j_a^b+ \gamma)J_a^t (j_a^b + \gamma)} =  \Delta_{j_a^b \frac{1}{2}j_a^{b \prime}}^2\Delta_{j_a^b J_a^t j_a^b} \Delta_{j_a^{b \prime} J_a^t j_a^{b \prime}} \ \ \ .
    \label{eq:4deltaEquality}
\end{equation}
Thus, the whole triangle delta part of the $6j$ symbol is invariant under the flux hierarchy inversion transformation (with the two latter triangle deltas swapping under the transformation).

Consider next the other half of the Racah $6j$ expansion given in Eq.~\eqref{eq: Racah 6j}, the sum over $J$,
\begin{equation}\label{eq: Racah sum over J}          \sum_{J=J_{\text{min}}}^{J_{\text{max}}}(-1)^{J}[J+1]!
    \left(\prod_{n = 1}^{3}[T_n - J]!\right)^{-1}\left(\prod_{m = 1}^{4}[J - \tau_{m}]!\right)^{-1},
\end{equation}
where $T$ is the set of tetrads and $\tau$ is the set of triads, as described in Appendix~\ref{sec: Appendix: F definition}. Prior to the transformation,
\begin{equation}
    J_\text{min} = \text{max}(\tau) = \text{max}\left(j_a^b + j_a^{b\prime} + 1/2, 2j_a^b + J_a^t, 2j_a^{b \prime} + J_a^t\right) \ \ \ .
\end{equation}
In the present context where $j_a^{b \prime} = j_a^b + 1/2$ for simplicity, $J_\text{max} = 2j_a^{b \prime} + J_a^t$. Similarly, because $2j_a^b \geq J_a^t$ by the triangle constraints, prior to the transformation
\begin{equation}
    J_\text{max} = \text{min}(T) = \text{min}\left( j_a^b + j_a^{b \prime} + 1/2 + J_a^t, 2j_a^b + 2j_a^{b \prime}\right) = 2j_a^{b \prime} + J_a^t = J_\text{min} \ \ \ .
\end{equation}
Thus, the sum over $J$ only has one term $J_\text{min} = J_\text{max} := J$. Now, after the transformation the upper bound becomes
\begin{equation}
        J_\text{min}' = \text{max}\left( j_a^b + j_a^{b \prime} + 2\gamma + 1/2, 2j_a^b + 2\gamma + J_a^t, 2j_a^{b \prime} + 2\gamma + J_a^t\right) = \text{max}\left( \tau + 2 \gamma\right) = 2\gamma + J \ \ \ ,
        \label{eq:Jminprime}
\end{equation}
where the triangle constraints apply as before. Similarly, the lower bound is 
\begin{equation}
        J_\text{max}' = \text{min}\left(j_a^b + j_a^{b \prime} + 2\gamma + 1/2 + J_a^t, 2j_a^b + 2j_a^{b \prime} + 4\gamma\right) = 2\gamma + J \ \ \ .
        \label{eq:Jmaxprime}
\end{equation}
 Thus, the single term in the sum is simply shifted upon flux hierarchy inversion, $J_\text{min}' = J_\text{max}' = 2\gamma + J := J'$.  
 As seen in Eq.~\eqref{eq:Jminprime}, the transformation shifts each of the triads by $2\gamma$, leading to the invariance of the triad factor in Eq.~\eqref{eq: Racah sum over J} due to cancellation with the $2\gamma$ shift in $J'$. 
 Of the two transformed tetrads, $j_a^b + j_a^{b \prime} + 2\gamma + J_a^t + 1/2$ and $2j_a^b + 2j_a^{b \prime} + 4\gamma$ (the former with multiplicity two), only the latter has a $\gamma$-shift that is not canceled by that in $J'$.
 With the $J$s appearing in $(-1)^J[J+1]!$ also shifted by $2\gamma$, all $\gamma$ dependence of the Racah formula is concentrated in the transformation of
\begin{equation}
    (-1)^{J}\frac{[J + 1]!}{[2j_a^b + 2j_a^{b \prime}  - J]!}
    \quad\rightarrow\quad
    (-1)^{J + 2\gamma}\frac{[J+2\gamma + 1]!}{[2j_a^b + 2j_a^{b \prime} + 2\gamma - J]!} \ \ \ .
\end{equation}
The $\gamma$-dependent factorial terms can be expanded
\begin{equation}
\label{eq: flux inversion J sum}
\begin{aligned}
    \frac{[J+2\gamma+1]!}{[2j_a^b + 2j_a^{b \prime} + 2\gamma - J]!} = \frac{[2j_a^{b\prime} + J_a^t + 2\gamma + 1]!}{[2j_a^b + 2\gamma - J_a^t]!} 
    &= [2j_a^b  + 2\gamma - J_a^t + 1]\ldots[2j_a^b  + J_a^t + 2\gamma + 2]  \\
    &= [k-J_a^t -2j_a^b]\ldots[k+1 +J_a^t -2j_a^b] \\
    &= [2j_a^b + J_a^t + 2]\ldots[2j_a^b - J_a^t + 1] \\
    &= \frac{[2j_a^{b \prime} + J_a^t + 1]!}{[2j_a^b - J_a^t]!} 
    = \frac{[J+1]!}{[2j_a^b + 2j_a^{b \prime} -J]!} \ \ \ ,
\end{aligned}
\end{equation}
indicating that the flux-hierarchy-inverted sum over $J'$ is the same as the original sum over $J$ up to a factor of $(-1)^{2\gamma}$. Because $\gamma$ can be either an integer or half-integer, this phase is non-vanishing.  Thus, the $6j$ factor within Eq.~\eqref{eq:fluxhierarchyinversionF} is invariant to flux hierarchy inversion up to a $(-1)^{2\gamma}$ phase.  

To complete analysis of the Eq.~\eqref{eq:fluxhierarchyinversionF} transformation,
in addition to the $6j$ coefficient, the phase and dimension factors within the F-symbol (Eq.~\eqref{eq: q-deformed F matrix}) must be considered.  
The former, $(-1)^{j_a^b + j_a^{b \prime} + 1/2 + J_a^t} \rightarrow (-1)^{j_a^b + j_a^{b \prime} + 2\gamma + 1/2 + J_a^t} $, contributes a second phase, accumulating to $(-1)^{4\gamma} = 1$ as $4\gamma \in 2\mathbb{N}_0$ is an even integer. The remaining $\gamma$ dependence is present only in the transformation of the dimension factors,
\begin{equation}
    \sqrt{D\left(j_a^{b \prime})D(j_a^b\right)}
    \quad \rightarrow \quad
    \sqrt{D\left(j_a^{b \prime} + \gamma\right)D\left(j_a^b + \gamma\right)}\ \ \ .
\end{equation}
Using the symmetry in Eq.~\eqref{eq: q-number under flux inversion}, these can be rewritten as
\begin{subequations}
    \begin{align}
    D(j_a^{b \prime}+\gamma) &= [2j_a^{b \prime} + 2\gamma + 1] = [-2j_a^{b \prime} + k + 2] = [2j_a^{b \prime}] = [2j_a^b + 1] = D(j_a^b)\\
    D(j_a^b+\gamma) &= [2j_a^b + 2\gamma + 1] = [-2j_a^b + k] = [2j_a^b + 2] = [2j_a^{b \prime} + 1] = D(j_a^{b \prime}) \ \ \ .
    \end{align}
\end{subequations}
As with the latter two triangle deltas in Eq.~\eqref{eq:4deltaEquality}, the flux hierarchy inversion transformation has swapped the arguments $j_a^b$ and $j_a^{b \prime}$ of the dimension factors.  Thus, the product $D(j_a^b)D(j_a^{b \prime})$ is invariant under the transformation. 

With the above, we have shown that the F-symbol in $\Box'''$ is invariant under the flux inversion transformation, i.e., the left and right sides of Eq.~\eqref{eq:fluxhierarchyinversionF} are equal. 
Finally, note that the additional phase associated with $\Box'''$ in Eq.~\eqref{eq: plaq after F3} is also invariant under the transformation, $\Delta j_a^b \rightarrow \left(j_a^{b \prime} + \gamma\right) - \left(j_a^b + \gamma\right) = \Delta j_a^b$. Thus, flux hierarchy inversion is a symmetry of $\Box'''$. 

This symmetry is unique to the q-deformed theory as the proof above requires the closure of the q-deformed factorial structure to hold. Under this symmetry, high-flux and low-flux states related by the flux hierarchy inversion transformation have the same matrix elements of the transformed plaquette operator $\Box'''$. In practice, this symmetry eliminates the need to calculate half the matrix elements of $\Box'''$.

\subsection{Enumerating actively mixed levels}\label{Appendix: counting actively mixed}
As introduced in Section~\ref{sec: Main: resource scaling}, reductions in resource scaling can be accomplished by recognizing that, for any control sector of a phased F-move or diagonalizing G-move, the associated unitary acts on an $m$-dimensional subspace of the full $d$-dimensional single-qudit space. Here, we provide formulae for the distribution of $m$-level unitaries for each part of the diagonalization sequence. In Appendix~\ref{sec: Appendix: decomp}, these formulae will be used to reduce circuit resources. 

The number of $m$-level unitaries required for a phased F-move controlled on four registers ($F_1$ and $F_2$) is
\begin{equation}
    n_4(m, k) = C_{k+3 - 2m}^{(4)} \ \ \ ,
\end{equation}
where $C_p^{(4)}$ are combinatorial factors defined by 
\begin{equation}
    C_p^{(4)} = \begin{cases} \delta_{p,1} + 8(p-1) + 16 \sum\limits_{\ell=1}^{p-1} \ell(p-1-\ell)
   = \delta_{p,1} + \frac{8}{3} \left( p^3 - 3p^2 + 5 p - 3\right), & p >0 \\
   0, &p\leq 0  \end{cases}\ \ \ .
   \label{eq:Cp4}
\end{equation}
Likewise, the number of $m$-level unitaries required for a phased F-move controlled on three registers is
\begin{equation}
    n_3(m, k) = C_{k+3 - 2m}^{(3)} \ \ \ ,
\end{equation}
where $C_p^{(3)}$ are combinatorial factors defined by 
\begin{equation}
    C_p^{(3)} 
    = \begin{cases}
      \delta_{p,1} + 1 + 3(2p-3) + 4 \sum\limits_{\ell=1}^{p-3} \ell 
      =
      -\delta_{p,1} + 2p^2 - 4p + 4, & p > 0\\
       0, & p \leq 0 \end{cases}  \ \ \ .
    \label{eq:Cp3}
\end{equation}
Explicit evaluation of the factors $C^{(4)}_p$ and $C^{(3)}_p$ relevant for $k \leq 8$ are provided in Table \ref{tab: level dist}.

\begin{center}
\begin{table}[h]
\begin{tabular}{ |c|c|c|c|c|c|c| } 
  \hline
    \multicolumn{7}{|c|}{F-move with four controls ($F_1, F_2$)} \\
\hline
 $k$ & $n_4(m=1, k)$ & $n_4(m=2, k)$ & $n_4(m=3, k)$ & $n_4(m=4, k)$ & $n_4(m=5, k)$ & $N_4(k)$ \\ 
 \hline
 $0$ & $C_1 = 1$ & 0 & 0 & 0 & 0 & 1 \\ 
 $1$ & $C_2 = 8$ & 0 & 0 & 0 & 0 & 8 \\ 
 $2$ & $C_3 = 32$ & $C_1$ & 0 & 0 & 0 & 33\\ 
 $3$ & $C_4 = 88$ & $C_2$ & 0 & 0 & 0 & 96\\
 $4$ & $C_5 = 192$ & $C_3$ & $C_1$ & 0 & 0 & 225\\
 $5$ & $C_6 = 360$ & $C_4$ & $C_2$ & 0 & 0 & 456\\
 $6$ & $C_7 = 608$ & $C_5$ & $C_3$ & $C_1$ & 0 & 833\\
 $7$ & $C_8 = 952$ & $C_6$ & $C_4$ & $C_2$ & 0 & 1408\\
 $8$ & $C_9 = 1408$ & $C_7$ & $C_5$ & $C_3$ & $C_1$ & 2241\\
\hline
\multicolumn{7}{|c|}{$N_4(k) = 1+\frac{8}{3}k + \frac{8}{3}k^2 + \frac{4}{3}k^3 + \frac{1}{3}k^4 \sim O(k^4)$}\\
\hline
\hline
    \multicolumn{7}{|c|}{F-move with three controls ($F_3$)} \\
\hline
 $k$ & $n_3(m=1, k)$ & $n_3(m=2, k)$ & $n_3(m=3, k)$ & $n_3(m=4, k)$ & $n_3(m=5, k)$ & $N_3(k)$ \\ 
 \hline
 $0$ & $C_1 = 1$ & 0 & 0 & 0 & 0 & 1 \\ 
 $1$ & $C_2 = 4$ & 0 & 0 & 0 & 0 & 4 \\ 
 $2$ & $C_3 = 10$ & $C_1$ & 0 & 0 & 0 & 11\\ 
 $3$ & $C_4 = 20$ & $C_2$ & 0 & 0 & 0 & 24\\
 $4$ & $C_5 = 34$ & $C_3$ & $C_1$ & 0 & 0 & 45\\
 $5$ & $C_6 = 52$ & $C_4$ & $C_2$ & 0 & 0 & 76\\
 $6$ & $C_7 = 74$ & $C_5$ & $C_3$ & $C_1$ & 0 & 119\\
 $7$ & $C_8 = 100$ & $C_6$ & $C_4$ & $C_2$ & 0 & 176\\
 $8$ & $C_9 = 130$ & $C_7$ & $C_5$ & $C_3$ & $C_1$ & 249\\
\hline
    \multicolumn{7}{|c|}{$N_3(k) = 1 + \frac{5}{3}k + k^2 + \frac{1}{3} k^3 \sim O(k^3)$}\\
\hline
\end{tabular}

\resizebox{\columnwidth}{!}{
\begin{tabular}{ |c|c|c|c|c|c|c|c|c|c| } 
\hline
    \multicolumn{10}{|c|}{Diagonalizing G} \\
\hline
 $k$ & $n_1(m=2, k)$ & $n_1(m=3, k)$ & $n_1(m=4, k)$ & $n_1(m=5, k)$ & $n_1(m=6, k)$ & $n_1(m=7, k)$ & $n_1(m=8, k)$ &$n_1(m=9, k)$ & $N_1(k)$ \\ 
 \hline
 0 & 0 & 0 & 0 & 0 & 0 & 0 & 0 & 0 & 0\\ 
 1 & 1 & 0 & 0 & 0 & 0 & 0 & 0 & 0 & 1\\ 
 2 & 0 & 1 & 0 & 0 & 0 & 0 & 0 & 0 & 1\\ 
 3 & 1 & 0 & 1 & 0 & 0 & 0 & 0 & 0 & 2\\
 4 & 0 & 1 & 0 & 1 & 0 & 0 & 0 & 0 & 2\\
 5 & 1 & 0 & 1 & 0 & 1 & 0 & 0 & 0 & 3\\
 6 & 0 & 1 & 0 & 1 & 0 & 1 & 0 & 0 & 3\\
 7 & 1 & 0 & 1 & 0 & 1 & 0 & 1 & 0 & 4\\
 8 & 0 & 1 & 0 & 1 & 0 & 1 & 0 & 1 & 4\\
\hline
    \multicolumn{10}{|c|}{$N_1(k) = \lceil \frac{k}{2}\rceil$}\\
\hline
\end{tabular}
}
\caption{Top and middle: The number of single-qudit unitaries mixing $m$ levels required to implement a (phased) F-move with four controls (top, Eq.~\eqref{eq:Cp4}) and three controls (middle, Eq.~\eqref{eq:Cp3}) for various truncation parameters $k$. Bottom: The number of $m$-level single qudit unitaries required to implement the diagonalizing G move or the transformed plaquette operator $\Box'''$, Eq.~\eqref{eq: n1}. The rightmost column in each case is the sum $N(k) = \sum_{m=1}^{k+1} n(m,k)$, which gives the total number of control sectors at a given truncation. Note that for $N_4$, $N_3$, the upperbound of the sum is $m = \lceil \frac{k+1}{2} \rceil$. To calculate corresponding circuit resources, we weight each $n(m,k)$ by a resource function $\xi_{O_m}$ as in Eq.~\eqref{eq: weighted depth sums}.}
\label{tab: level dist}
\end{table}
\end{center}

The structure of the diagonalizing G unitaries is dictated by the structure of the transformed plaquette operator $\Box''' = F_3 F_2 F_1 \Box F_1^\dagger F_2^\dagger F_3^\dagger$ that they diagonalize. After the three phased F-moves, the transformed plaquette operator in Eq.~\eqref{eq: plaq after F3} vanishes for all non-integer values of $J_a^t$ because the relevant F-symbol,
$
\begin{bmatrix}
    j_a^b & 1/2 & j_a^{b \prime} \\
    j_a^{b \prime} & J_a^t & j_a^b
\end{bmatrix}
$,
enforces the integer-sum gauge singlet constraints $J_a^t + 2j_a^{b(\prime)} \in \mathbb{N}_0$ for both primed and unprimed $j_a^b$. Intuitively, this requirement can also be read off of the modified lattice diagram in Fig.~\ref{fig: f sequence} whose physical flux configurations require that any flux through the $j_a^b$ link flows through the $J_a^t$ link, disallowing half-integer flux on $J_a^t$. So, there are $\lceil \frac{k+1}{2} \rceil$ non-trivial control sectors, $J_a^t$, of $\Box''' = \prod_{J_a^t}\Box'''(J_a^t)$. The triangle inequalities enforced by the F-symbol in Eq.~\eqref{eq: plaq after F3} additionally constrain the transitions $|j_a^b\rangle \rightarrow |j_a^{b \prime}\rangle$ allowed by the transformed plaquette operator such that $|j_a^b - j_a^{b \prime}| = 1/2$. For each valid value of $J_a^t \in \{0, 1, \ldots, \lfloor k/2\rfloor \}$, there are as many transitions as there are adjacent-level pairs ($j_a^b, j_a^{b \prime} = j_a^b \pm 1/2$) satisfying the fusion and singlet  constraints $2j_a^{b(\prime)} + J_a^t \leq k$ and $2j_a^b \geq J_a^t$. For fixed $k$ and $J_a^t$, there are $m = k + 1 - 2J_a^t$ valid initial $j_a^b$ corresponding to $m-1$ interactions between adjacent levels. For control sectors $J_a^t$ with $m = 1$, the transformed plaquette operator vanishes (as no transitions satisfy the fusion constraint), requiring no quantum circuitry for implementation.

The G unitaries $G(J_a^t)$ share the structure of $\Box'''(J_a^t)$, so that for each  control sector $J_a^t$ of the transformed plaquette operator the G unitary that diagonalizes this operator is likewise an $m$-level operator. Thus, for fixed $k$ the transformed plaquette operator (and the corresponding G unitaries) requires one $m$-level unitary for each $m \in \{2, 4, \ldots, k+1\}$ for odd $k$ or $m \in \{3, 5, \dots, k+1\}$ for even $k$. That is,
\begin{equation}\label{eq: n1}
    n_{\Box^{'''}}(m,k) = n_G(m,k) =
    \begin{cases}
        1, \text{ $2 \leq m \leq k+1$, $m$ even, $k$ odd} \\
        1, \text{ $3 \leq m \leq k+1$, $m$ odd, $k$ even}
        \\
        0, \text{ else}
    \end{cases}  \equiv n_1(m,k)\ \ \ .
\end{equation}

\section{Circuit implementation}\label{sec: Appendix: circuit implementation}
\subsection{General prescription for circuits}\label{sec: Appendix: recipe}
In this section, a procedure is detailed for determining multi-controlled unitaries capable of implementing our GVC completions that capture the phased F-moves in the gauge invariant space. In Appendix~\ref{sec: Appendix: decomp}, we will analyze a basic circuit synthesis strategy for these multi-controlled unitaries.  The procedure outlined here prepares for this analysis in connection with our diagrammatic approach to the phased F-moves.

\begin{enumerate}
    \item Determine control sectors $\{a,b,c,d\}$ corresponding to physical flux configurations of the original lattice. The total number of control sectors that need to be accounted for is given in the $N(k)$ column of Table \ref{tab: level dist}. Note that some valid control sectors of the non-deformed theory are invalid under the fusion-constraint of the q-deformed theory.

    \begin{quote} \textbf{Ex:} for $k = 3$, $\{a,b,c,d\} = \{1,1,3/2,3/2\}$ is a valid control sector with or without q-deformation. However, $\{a,b,c,d\} = \{0,1,3/2,3/2\}$ is not a valid control sector of the q-deformed theory despite being a valid control sector of the non-deformed theory.
    \end{quote}
    
    \item For each control sector, identify physical flux configurations of the modified lattice with the same controls (external links). Assign the transition amplitude $\langle a,b,c,d,j'|F_i|a,b,c,d,j\rangle$ specified by the phased F-move to the  $|j'\rangle \langle j|$ element of the single-qudit matrix acting on the $|j\rangle$ register.

    \begin{quote}
        \textbf{Ex:} for $k = 3$ and controls $\{a,b,c,d\} = \{1,1,3/2,3/2\}$, only the following F-move transition is allowed:
        \begin{center}
        \includegraphics[width=0.35\linewidth]{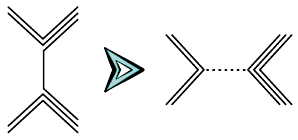}
        \end{center}
        Thus, the phased $F_1$ has the following matrix representation in the computational basis
        $$F_1(1,1,3/2,3/2) = \begin{bmatrix}
            0 & -i & 0 & 0 \\
            0 & 0 & 0 & 0 \\
            0 & 0 & 0 & 0 \\
            0 & 0 & 0 & 0
        \end{bmatrix}.$$
    \end{quote}
    
    \item For a given control sector, if $\mathcal{H}_{\text{phys}}  = \mathcal{H}_{\text{phys}}'$, the non-vanishing matrix elements of the phased F-move in the single-qudit active space will be located within a diagonal block, not necessarily of consecutive basis states. The unitary matrix can be completed by introducing identity over all states in the unphysical subspace.

    \begin{quote}
        \textbf{Ex:} for $k = 3$ and controls $\{a,b,c,d\} = \{1,1/2,1,1/2\}$, the $|j\rangle$ subspace of $\mathcal{H}_\text{phys}$ and $\mathcal{H}_\text{phys}'$ is $\{|1/2\rangle, |3/2\rangle\}$. We thus have the following phased F-move, centered on the diagonal,
        
        $$
        F_1(1,1/2,1,1/2) = 
        \begin{bmatrix}
            \textcolor{teal}{0} & 0 & 0 & 0 \\
            0 & a & 0 & b \\
            0 & 0 & \textcolor{teal}{0} & 0 \\
            0 & b & 0 & -a
        \end{bmatrix}$$
    
    with $a = \frac{\sqrt{5}-1}{2}$ and $b = \left( \frac{3-\sqrt{5}}{2} \right)^{1/4}$. A unitary completion with identity over $\mathcal{H}_\text{unphys}$ would be $U_{F_1}(1,1/2,1,1/2) = F_1\left(1,1/2,1,1/2\right) + |0\rangle\langle0| + |1\rangle\langle1|$, replacing the highlighted elements with unity.
    \end{quote}
    \item Else, if $\mathcal{H}_{\text{phys}} \neq \mathcal{H}_{\text{phys}}'$ the non-vanishing matrix elements of the phased F-move will be clustered in a (generally non-consecutive) off-diagonal block. A unitary completion can be furnished by hand, if an obvious choice is apparent. In general a systematic approach is to apply a unitary Pauli $X$ (or series thereof) to center the block along the diagonal. Then, the identity completion strategy of step 3 can be implemented, which now introduces $\mathcal{H}_{\text{unphys}} \rightarrow \mathcal{H}_{\text{unphys}}'$ transitions.

        \begin{quote}
            \textbf{Ex:} for $k = 3$ and controls $\{a,b,c,d\} = \{1,1,1/2,1/2\}$, the $|j\rangle$ subspace of $\mathcal{H}_\text{phys}$ is $\{|1/2\rangle, |3/2\rangle\}$ and that of $\mathcal{H}_\text{phys}'$ is $\{|0\rangle, |1\rangle\}$. The associated phased F-move is thus not centered on the diagonal
        $$
        F_1(1,1,1/2,1/2) = 
        \begin{bmatrix}
            0 & ib & 0 & ia \\
            \textcolor{teal}{0} & 0 & 0 & 0 \\
            0 & ia & 0 & -ib \\
            0 & 0 & \textcolor{teal}{0} & 0
        \end{bmatrix}.$$
        To determine a unitary completion, we may apply two-level Pauli $X$ operators to center the non-zero block,
        $$ U_{F_1}(1,1,1/2,1/2) =   
        \begin{bmatrix}
            ib & 0 & ia & 0 \\
            0 & \textcolor{teal}{1} & 0 & 0 \\
            ia & 0 & -ib & 0 \\
            0 & 0 & 0 & \textcolor{teal}{1}
        \end{bmatrix}X_{01}X_{23} \ \ \ ,
        $$
        where identities have been placed along the highlighted diagonals as in step 3.
        \end{quote}
\end{enumerate}

The diagonalizing G operators are found using a similar procedure. For each value of the $|J_a^t\rangle$ register that $G(J_a^t)$ is controlled on, we write the Hermitian matrix representing the transformed plaquette operator for that control value $\Box'''(J_a^t)$ as given by Eq.~\eqref{eq: plaq after F3} in the single-qudit $|j_a^b\rangle$ subspace. Since this subspace has dimensions $d \times d$ (with $d = k+1$), the unitary that diagonalizes this matrix can be calculated for local gauge field truncations anticipated for quantum simulation applications.

\subsection{Qubit and qutrit circuits}\label{Appendix: circuits}
\begin{figure}
    \centering
    \includegraphics[width=0.5\linewidth]{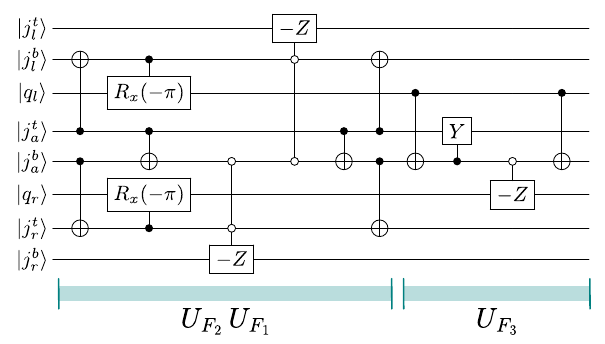}
    \caption{An alternative circuit capable of implementing the phased F-moves at qubit truncation.  By computing intermediate parities, the total GCX cost of the plaquette Trotter step can be reduced to an upperbound of 48.}
    \label{fig: reduced parity circuit}
\end{figure}

Following the procedure outlined in the main text and Appendix~\ref{sec: Appendix: recipe}, circuits implementing the phased F-moves for $k = 1$ and $k = 2$ truncations are presented in Tables~\ref{tab: qubit elements} and~\ref{tab: qutrit elements}, respectively. 
For the qubit truncation, choosing an alternative GVC (relative to that in Fig.~\ref{fig: schematic circuit}) enables each of the $\mathcal{H}_\text{phys} \rightarrow \mathcal{H}_\text{phys}'$ transitions to be implemented as $R_x$ rotations with the same rotation angle for off-diagonal transitions (see the rightmost column of Table~\ref{tab: qubit elements}). 
Rather than employing uniformly controlled techniques~\cite{Di:2013qvb,Jiang:2025ufg}, we here report further reductions to the GCX cost by controlling F-move operators on strategically computed parities, as shown in Fig.~\ref{fig: reduced parity circuit}. 

Recognizing that off-diagonal transitions of $U_{F_1}$ and $U_{F_2}$ correspond to states with odd $2(b+d)$ and that diagonal transitions have even $2(b+d)$, an $R_x(-\pi)$ implementing the off-diagonal transitions can be acted over all $2(b+d)$-odd control sectors. With indices relating to Table~\ref{tab: qubit elements}, $\{b, d\}$ are registers $\{|j_l^b\rangle, |j_a^t\rangle\}$ for $U_{F_1}$ and are registers $\{|j_a^b\rangle, |j_r^t\rangle\}$ for $U_{F_2}$. 
Note that this realizes extra rotations in the GVC over the unphysical sectors with odd $2(b+d)$. 
The one non-trivial diagonal transition of the $2(b+d)$-even sectors must be further specified by calculation of an additional parity. Since $2(j_a^t+j_a^b)$ serves this purpose for both $U_{F_1}$ and $U_{F_2}$, this parity only needs to be calculated once. 
By applying a $-Z$ gate in the sector of both even parities, the physical $\{0,0,0,0\}$ and unphysical $\{1,1,1,0\}$ control sector experiences $(-1)$ phase.  
These parities are then uncomputed to complete the $U_{F_2}U_{F_1}$ implementation.

By a similar logic for $U_{F_3}$, the parity of $2(a+c) = 2(q_l + j_a^b)$ differentiates the off-diagonal transitions (odd) and the diagonal transitions (even). The $U_{F_3}$ transformation can thus be accomplished by computing this parity and controlling the $Y$($-Z$) gates on the odd(even) parity. 
With the $-\pi$ rotation angle allowing $CR_x(-\pi) = (S\otimes \mathbb{I})CNOT$,  each single-controlled gate of Fig.~\ref{fig: reduced parity circuit} requires one GCX.
Meanwhile, the double-controlled $-Z$ gate requires six via the standard optimal Toffoli decomposition~\cite{Shende:2008plv}.
Thus, the combined parity- and GVC-based reductions described here achieve a total GCX cost of 48 for one Trotter step of the plaquette time evolution.

\begin{table}[h!]\label{tab: qubit elements}
\centering
        \centering
        \begin{tabular}{|c|c|c|c|c|c|}
        \hline
        \multicolumn{6}{|c|}
        {Four-controlled phased F-moves ($F_1, F_2$)}\\
        \hline
        \begin{tabular}{c} Controls ~\\ 2\{a,b,c,d\} \end{tabular} & \begin{tabular}{c} Transition ~\\ $|j\rangle \rightarrow |J\rangle$ \end{tabular} & \begin{tabular}{c} Amplitude ~\\ $\langle J|F_{1,2}|j\rangle$ \end{tabular} & Gate & \begin{tabular}{c}Fig.~\ref{fig: schematic circuit} gate \\ (with global phase) \end{tabular} & \begin{tabular}{c}Fig.~\ref{fig: reduced parity circuit} gate \\ (with global phase \\ and alternative GVC) \end{tabular} \\
        \hline \hline
        \{0,0,0,0\} & $|0\rangle \rightarrow |0\rangle$  & $1$                                                        & $\mathbb{I}$  & $-\mathbb{I} \rightarrow C^{(3)}(-Z)$ & $C^{(3)}(-Z)$                   \\
        \hline
        
        \{0,0,1,1\} & $|1/2\rangle \rightarrow |0\rangle$    & $-i$                                                        & $Y$ & $-Y$ & $R_x(-\pi)$                                   \\
        \hline
        
        \{0,1,0,1\} & $|1/2\rangle \rightarrow |1/2\rangle$      & $-1$                                                        & $-\mathbb{I}$  & $\mathbb{I}$ &  $\mathbb{I}$                                 \\
        \hline
        
        \{0,1,1,0\} & $|0\rangle \rightarrow |1/2\rangle$    & $-i$                                                        & $-Y$ & $Y$ & $R_x(-\pi)$                                   \\
        \hline
        
        \{1,0,0,1\} & $|0\rangle \rightarrow |1/2\rangle$      & $-i$                                                        & $-Y$ & $Y$ & $R_x(-\pi)$                                  \\
        \hline
        
        \{1,0,1,0\} & $|1/2\rangle \rightarrow |1/2\rangle$    & $-1$                                                         & $-\mathbb{I}$ & $\mathbb{I}$ & $\mathbb{I}$                                \\
        \hline
        
        \{1,1,0,0\} & $|1/2\rangle \rightarrow |0\rangle$      & $-i$                                                         & $Y$ & $-Y$ & $R_x(-\pi)$                           \\
        \hline

        \{1,1,1,1\} & $|0\rangle \rightarrow |0\rangle$  & $-1$                                                        & $-\mathbb{I}$  & $\mathbb{I} \rightarrow C^{(3)}(-Z)$  & $C^{(3)}(-Z)$                  \\
        \hline    
        \end{tabular}

    \bigskip
    
        \centering
        \begin{tabular}{|c|c|c|c|c|}
        \hline
        \multicolumn{5}{|c|}
        {Three-controlled phased F-moves ($F_3$)}\\
        \hline
        \begin{tabular}{c} Controls ~\\ 2\{a,c,d\} \end{tabular} & \begin{tabular}{c} Transition ~\\ $|j\rangle \rightarrow |J\rangle$ \end{tabular} & \begin{tabular}{c} Amplitude ~\\ $\langle J|F_3|j\rangle$ \end{tabular} & Gate & Fig.~\ref{fig: schematic circuit} gate \\
        \hline \hline
        
        \{0,0,0\} & $|0\rangle \rightarrow |0\rangle$  & $1$ & $\mathbb{I}$  & $-\mathbb{I} \rightarrow C^{(2)}(-Z)$                \\
        \hline
        \{0,1,1\} & $|1/2\rangle \rightarrow |0\rangle$    & $i$ & $-Y$ & $Y$                                \\
        \hline
        
        \{1,0,0\} & $|1/2\rangle \rightarrow |0\rangle$      & $i$  & $-Y$  & $Y$                          \\
        \hline
        \{1,1,1\} & $|0\rangle \rightarrow |0\rangle$  & $-1$   &  $-\mathbb{I}$ & $\mathbb{I}$                  \\
        \hline
        \end{tabular} 

    \bigskip
    
        \centering
        \begin{tabular}{|c|c|c|c|}
        \hline
        \multicolumn{4}{|c|}
        {G-move}\\
        \hline
        Control $|J_a^t\rangle$ & $\Box'''(J_a^t)$ & $G(J_a^t)$ & \begin{tabular}{c} Diagonal plaq. \\ $\tilde{\Box}(J_a^t)$ \end{tabular} \\
        \hline
        \hline
        0 & $\begin{bmatrix} 
            0 & i\\
            -i & 0
            \end{bmatrix}$ 
            & $HS$ & $Z$               \\
        \hline
        \end{tabular}
\caption{Qubit ($k = 1$) circuitry content of phased F-moves organized by control sector. For each control sector, the non-trivial transition(s) and their  amplitudes under the phased F-moves are given, along with unitary gates that implement the  prescribed transitions with a chosen unitary GVC over the unphysical space. Top: four-controlled phased F-moves of the form $\langle J|F|j\rangle = (-1)^{-(j+J)}$\scalebox{0.75}{$\begin{bmatrix} a&b&J\\ c&d&j\end{bmatrix}$} as in $F_1$, $F_2$. The second-from-right column shows the Pauli gate completion employed in Fig.~\ref{fig: schematic circuit}. The rightmost column provides an alternative completion using $R_x(\theta) = \exp(-i\frac{X}{2}\theta)$ suitable for uniformly controlled techniques and for the circuit in Fig.~\ref{fig: reduced parity circuit}. Middle: three-controlled F-moves of the form $\langle J|F|j\rangle = (-1)^{j+J}$\scalebox{0.75}{$\begin{bmatrix} a&a&J\\ c&d&j\end{bmatrix}$} as in $F_3$. Bottom: G-move that diagonalizes the transformed plaquette operator $\Box''' = F_3 F_2 F_1 \Box F_1^\dagger F_2^\dagger F_3^\dagger$ defined in Eq. (\ref{eq: plaq after F3}).}
\label{tab:quditF1F2F3Gunitaries}
\end{table}

\begin{table}[h!]\label{tab: qutrit elements}
\begin{minipage}{0.48\textwidth}
    \begin{tabular}{|c|c|c|c|}
        \hline
        \multicolumn{4}{|c|}
        {Four-controlled phased F-moves ($F_1, F_2$)}\\
        \hline
         2\{a,b,c,d\}  & $|j\rangle \rightarrow |J\rangle$  &   $\langle J|F_{1,2}|j\rangle$  & Gate\\
        \hline \hline
        \{0,0,0,0\} & $|0\rangle \rightarrow |0\rangle$      & 1                                                         & $\mathbb{I}$                               \\
        \hline
        \{0,0,1,1\} & $|1/2\rangle \rightarrow |0\rangle$    & $-i$                                                        & $Y_{01}$                                      \\
        \hline
        \{0,0,2,2\} & $|1\rangle \rightarrow |0\rangle$      & $-1$                                                        & $-X_{02}$                                     \\
        
        \hline
        \{0,1,0,1\} & $|1/2\rangle \rightarrow |1/2\rangle$  & $-1$                                                        & $-\mathbb{I}$                       \\
        \hline
        
        \{0,1,1,0\} & $|0\rangle \rightarrow |1/2\rangle$    & $-i$                                                        & $-Y_{01}$                                     \\
        \hline
        
        \{0,1,1,2\} & $|1\rangle \rightarrow |1/2\rangle$    & $i$                                                         & $-Y_{12}$                                     \\
        \hline
        
        \{0,1,2,1\} & $|1/2\rangle \rightarrow |1/2\rangle$  & $-1$                                                        & $-\mathbb{I}$                       \\
        \hline
        
        \{0,2,0,2\} & $|1\rangle \rightarrow |1\rangle$      & 1                                                         & $\mathbb{I}$                               \\
        \hline
        
        \{0,2,1,1\} & $|1/2\rangle \rightarrow |1\rangle$    & $i$                                                         & $Y_{12}$                                      \\
        \hline
        
        \{0,2,2,0\} & $|0\rangle \rightarrow |1\rangle$      & $-1$                                                        & $-X_{02}$                                     \\
        \hline
        
        \{1,0,0,1\} & $|0\rangle \rightarrow |1/2\rangle$    & $-i$                                                        & $-Y_{01}$                                     \\
        \hline
        \{1,0,1,0\} & $|1/2\rangle \rightarrow |1/2\rangle$  & $-1$                                                        & $-\mathbb{I}$                       \\
        \hline
        \{1,0,1,2\} & $|1/2\rangle \rightarrow |1/2\rangle$  & $-1$                                                        & $-\mathbb{I}$                       \\
        \hline
        
        \{1,0,2,1\} & $|1\rangle \rightarrow |1/2\rangle$    & $i$                                                         & $-Y_{12}$                                     \\
        \hline
        
        \{1,1,0,0\} & $|1/2\rangle \rightarrow |0\rangle$    & $-i$                                                        & $Y_{01}$                                      \\
        \hline
        
        \{1,1,0,2\} & $|1/2\rangle \rightarrow |1\rangle$    & $i$                                                         & $Y_{12}$                                      \\
        \hline
        
        \{1,1,1,1\} & $|0\rangle, |1\rangle \rightarrow |0\rangle,|1\rangle$      & $(-1)^{jJ+1}/\sqrt{2}$                                                & $-H_{02}$          \\
        \hline

        \{1,1,2,0\} & $|1/2\rangle \rightarrow |1\rangle$    & $i$                                                         & $Y_{12}$                                      \\
        \hline
        
        \{1,1,2,2\} & $|1/2\rangle \rightarrow |0\rangle$    & $i$                                                         & $-Y_{01}$                                     \\
        \hline
        
        \{1,2,0,1\} & $|1\rangle \rightarrow |1/2\rangle$    & $i$                                                         & $-Y_{12}$                                     \\
        \hline
        
        \{1,2,1,0\} & $|1/2\rangle \rightarrow |1/2\rangle$  & $-1$                                                        & $-\mathbb{I}$                      \\
        \hline
        
        \{1,2,1,2\} & $|1/2\rangle \rightarrow |1/2\rangle$  & 1                                                         & $\mathbb{I}$                               \\
        \hline
        
        \{1,2,2,1\} & $|0\rangle \rightarrow |1/2\rangle$    & $i$                                                         & $Y_{01}$                                      \\
        \hline
        
        \{2,0,0,2\} & $|0\rangle \rightarrow |1\rangle$      & $-1$                                                        & $-X_{02}$                                     \\
        \hline
        
        \{2,0,1,1\} & $|1/2\rangle \rightarrow |1\rangle$    & $i$                                                         & $Y_{12}$                                      \\
        \hline
        
        \{2,0,2,0\} & $|1\rangle \rightarrow |1\rangle$      & 1                                                         & $\mathbb{I}$                              \\
        \hline
        
        \{2,1,0,1\} & $|1/2\rangle \rightarrow |1/2\rangle$  & $-1$                                                        & $-\mathbb{I}$                      \\
        \hline
        \{2,1,1,0\} & $|1\rangle \rightarrow |1/2\rangle$    & $i$                                                         & $-Y_{12}$                                     \\
        \hline
        
        \{2,1,1,2\} & $|0\rangle \rightarrow |1/2\rangle$    & $i$                                                         & $Y_{01}$                                      \\
        \hline
        
        \{2,1,2,1\} & $|1/2\rangle \rightarrow |1/2\rangle$  & 1                                                         & $\mathbb{I}$                               \\
        \hline
        
        \{2,2,0,0\} & $|1\rangle \rightarrow |0\rangle$      & $-1$                                                        & $-X_{02}$ \\  
        \hline
        
        \{2,2,1,1\} & $|1/2\rangle \rightarrow |0\rangle$    & $i$                                                         & $-Y_{01}$                                     \\
        \hline
        
        \{2,2,2,2\} & $|0\rangle \rightarrow |0\rangle$      & 1                                                         & $\mathbb{I}$                              \\
        \hline
    \end{tabular}
    \end{minipage}
    \begin{minipage}{0.47\textwidth}
    \begin{tabular}{|c|c|c|c|}
        \hline
         \multicolumn{4}{|c|}{Three-controlled phased F-move ($F_3$)}\\
        \hline
        \begin{tabular}{c} Controls ~\\ 2\{a,c,d\} \end{tabular} & \begin{tabular}{c} Transition ~\\ $|j\rangle \rightarrow |J\rangle$ \end{tabular} & $\langle J|F_3|j\rangle$ & Gate\\
        \hline \hline
        \{0,0,0\} & $|0\rangle \rightarrow |0\rangle$      & $1$                                                         & $\mathbb{I}$                               \\
        \hline
        
        \{0,1,1\} & $|1/2\rangle \rightarrow |0\rangle$      & $i$                                                         & $-Y_{01}$                               \\
        \hline
        \{0,2,2\} & $|1\rangle \rightarrow |0\rangle$    & $-1$                                                       & $-X_{02}$                                      \\
        \hline
        
        \{1,0,0\} & $|1/2\rangle \rightarrow |0\rangle$      & $i$                                                        & $-Y_{01}$                                     \\
        \hline
        
        \{1,0,2\} & $|1/2\rangle \rightarrow |1\rangle$  & $-i$                                                        & $-Y_{12}$                       \\
        \hline
        
        \{1,1,1\} & $|0\rangle \rightarrow |0\rangle$      & $-1/\sqrt{2}$                                                & $-H_{02}$          \\
                    & $|1\rangle \rightarrow |1\rangle$~     & $1/\sqrt{2}$~                                                &                                          \\
                    & $|1\rangle \rightarrow |0\rangle$      & $-1/\sqrt{2}$                                                &                                          \\
                    & $|0\rangle \rightarrow |1\rangle$      & $-1/\sqrt{2}$                                                &                                          \\
        \hline
        
        \{1,2,0\} & $|1/2\rangle \rightarrow |1\rangle$    & $-i$                                                        & $-Y_{12}$                                     \\
        
        \hline
        \{1,2,2\} & $|1/2\rangle \rightarrow |0\rangle$  & $-i$                                                        & $Y_{01}$                       \\
        \hline
        
        \{2,0,0\} & $|1\rangle \rightarrow |0\rangle$      & $-1$                                                        & $-X_{02}$                                     \\
        \hline
        
        \{2,1,1\} & $|1/2\rangle \rightarrow |0\rangle$    & $-i$                                                         & $Y_{01}$                                     \\
        \hline
        
        \{2,2,2\} & $|0\rangle \rightarrow |0\rangle$      & $1$                                                         & $\mathbb{I}$                               \\
        \hline
    \end{tabular}
    
    \vspace{1cm}
    
    \begin{tabular}{|c|c|c|c|}
        \hline
        \multicolumn{4}{|c|}
        {G-move}\\
        \hline
        \begin{tabular}{c}Control \\ $|J_a^t\rangle$ \end{tabular} & $\Box'''(J_a^t)$ & $G(J_a^t)$ &  $\tilde{\Box}(J_a^t)$ \\
        \hline 
        \hline
        0 & $\scalemath{0.75}{\begin{bmatrix} 
            0 & i & 0 \\
            -i & 0 & i \\
            0 & -i & 0
            \end{bmatrix}}$  & 
            $S_1 e^{i\frac{\pi}{2} S_x}$ 
            & $\scalemath{0.75}{\begin{bmatrix} 
            \sqrt{2} & & \\
            & 0 & \\
            & &-\sqrt{2}
            \end{bmatrix}}$               \\
        \hline
    \end{tabular}
    \end{minipage}
\caption{Qutrit ($k = 2$) circuitry content of phased F-moves organized by control sector. For each control sector $\{ \cdot\}$, the non-trivial $|j\rangle \rightarrow |J\rangle$ transition(s) and their  amplitudes under the phased F-moves are given, along with unitary gates that implement the prescribed transitions with a chosen unitary GVC over the unphysical space. Left: four-controlled phased F-moves of the form $\langle J|F|j\rangle = (-1)^{-(j+J)}$\scalebox{0.75}{$\begin{bmatrix} a&b&J\\ c&d&j\end{bmatrix}$} as in $F_1$, $F_2$. Right: three-controlled F-moves of the form $\langle J|F|j\rangle = (-1)^{j+J}$\scalebox{0.75}{$\begin{bmatrix} a&a&J\\ c&d&j\end{bmatrix}$} as in $F_3$ and G-move diagonalizing the transformed plaquette operator of Eq.~\eqref{eq: plaq after F3}.}
\end{table}

\FloatBarrier
\subsection{Calculation of circuit resource scalings}
\label{sec: Appendix: decomp}

A naïve expression for the circuit resources of a plaquette operator Trotter step is given in Eq.~\eqref{eq: Trotter depth}. However, as discussed in Appendix~\ref{Appendix: counting actively mixed}, we are able to reduce this scaling by 1.) recognizing that the unitaries that constitute the phased F-moves act on a subspace of the single qudit Hilbert space and 2.) performing time evolution under $\Box'''$ by interleaving control sectors so that the diagonalizing G unitaries $G(J_a^t)$ do not need to be controlled. Further reductions to GCX resource scaling are afforded by tailoring our decomposition scheme to take advantage of GVC freedom and the anti-symmetry of $\Box'''$.

Before detailing the circuit synthesis scheme, let us establish two alterations to the circuit resources expression, Eq.~\eqref{eq: Trotter depth}. As described in Appendix~\ref{Appendix: counting actively mixed}, the multi-controlled unitaries associated with the (phased) F-moves each mix only $m \leq \lceil\frac{k+1}{2}\rceil < d$ levels of the $d$-dimensional single-qudit Hilbert space (with the value of $m$ being control-sector dependent). Matrix elements among the remaining $d - m$ levels correspond to unphysical to unphysical transitions, and we choose a GVC that acts as the identity over that subspace after centering (producing $\mathcal{H}_{\text{phys}} = \mathcal{H}_{\text{phys}}'$), as described in Appendix~\ref{sec: Appendix: recipe}. Here, we use the level distributions provided in Appendix~\ref{Appendix: counting actively mixed} to assign the appropriate depth and number of GCX gates to each controlled unitary of the circuit.  At worst, centering requires $m$ two-level $X_{ij}$ gates, which can be implemented in depth $\mathcal{O}(m)$ or, if operations on distinct levels can be parallelized as in \cite{Clements:2016rkw}, depth $\mathcal{O}(1)$. Naturally, quantifying the resource scaling of GCX gates is not affected by this choice.

Let $\xi_O$ be the circuit resources of operator $O$. Then, the cost of each phased F-move can be written as a weighted sum over the cost of $m$-level unitaries
\begin{equation}
    \begin{gathered}
    \xi_{F_{1,2}}(k) = \sum_{m=1}^{\lceil\frac{k+1}{2}\rceil} n_4(m,k) \xi_{C^{(4)}U_m}
    \\
    \xi_{F_{3}}(k) = \sum_{m=1}^{\lceil\frac{k+1}{2}\rceil} n_3(m,k) \xi_{C^{(3)}U_m}
    \\
    \xi_G(k) = \sum_{m=2}^{k+1} n_1(m,k) \xi_{CU_m} \ \ \ ,
    \end{gathered}
    \label{eq: weighted depth sums}
\end{equation}
where the $n(m,k)$ are the number of controlled unitaries associated with the operators discussed in Appendix~\ref{Appendix: counting actively mixed}.

\begin{figure}
    \centering
    \includegraphics[width=0.6\linewidth]{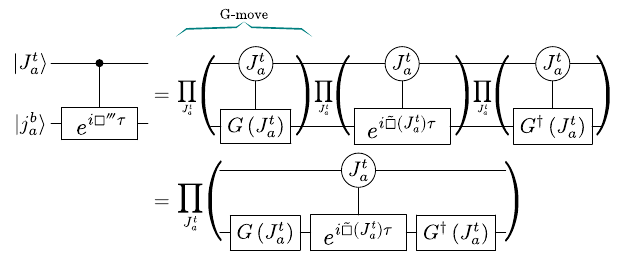}
    \caption{The controlled G-move can be decomposed into uncontrolled single-qudit operators if the transformed plaquette operator is performed in steps for each control sector ($J_a^t$). When utilized for Trotterized evolution of the magnetic term, $\tau = \frac{t}{g^2N_T}$. While the choice between these two strategies may depend on the parallelization capabilities of particular quantum architectures, the latter reduces basic GCX resources.}
    \label{fig: gmove to gu}
\end{figure}

Depending on the specifics of a quantum architecture, it may be advantageous to perform the controlled G-move and diagonalized plaquette time evolution interleaved control-sector-by-control-sector, as shown in Fig.~\ref{fig: gmove to gu}. With this approach, each $G(J_a^t)$ does not need to be controlled explicitly on the $|J_a^t\rangle$ register. Instead of $\xi_G$, we will consider the resource scaling of the transformed plaquette operator, $\xi_{\Box'''}$, where for each control sector $\Box'''(J_a^t) = G^\dagger(J_a^t)\tilde{\Box}(J_a^t)G(J_a^t)$ such that the diagonalizing G unitaries are non-controlled operators acting on the target register of the controlled diagonal time evolution operators $\tilde{\Box}(J_a^t)$. In this case, instead of Eq.~\eqref{eq: Trotter depth}, we evaluate
\begin{equation}\label{eq: modified Trotter depth with gu}
    \xi_{\text{Trot}} = 4 \xi_{F_{1,2}} + 2\xi_{F_3} + \xi_{\Box'''},
\end{equation}
where $\xi_{F_{1,2}}, \xi_{F_3}$ are as in Eq.~\eqref{eq: weighted depth sums} and $\xi_{\Box'''} =\sum_m n_1(m,k)\xi_{CU_m}$ is the resource cost of the whole $\Box'''$ evolution with interleaved control sectors modulo single-qudit rotations.

Having written an alternative expression of the circuit resources that recognizes the number of levels that each circuit component acts on, we decompose these components into elementary two-level single-qudit rotations (Givens) and generalized controlled-X (GCX) gates. A scheme for decomposing each multi-controlled unitary into this set of gates is presented below and summarized in Table~\ref{tab:gcxCalc}.

\begin{figure}[h!]
    \centering
    \includegraphics[width=0.75\linewidth]{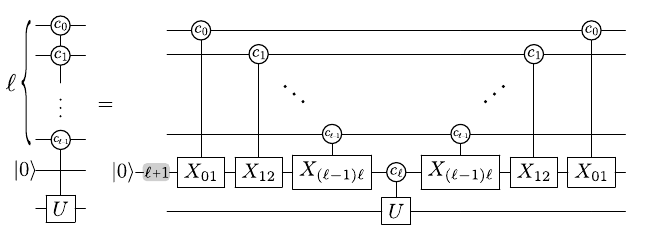}
    \caption{Decomposition of a single-qudit unitary controlled on $\ell$ registers, using GCX gates acting on an auxiliary qudit of dimension $\ell+1$.}
    \label{fig: multicontrol decomp}
\end{figure}

First, multi-controlled operators can be decomposed into single-controlled operators. Using an auxiliary qudit of dimension $\ell+1$, each single-qudit unitary controlled on $\ell$ registers $\text{C}^{(\ell)}\text{U}$ can be decomposed into $2\ell$ GCX gates and one CU gate as shown in Fig.~\ref{fig: multicontrol decomp}.

\begin{figure}
    \centering
    \includegraphics[width=0.8\linewidth]{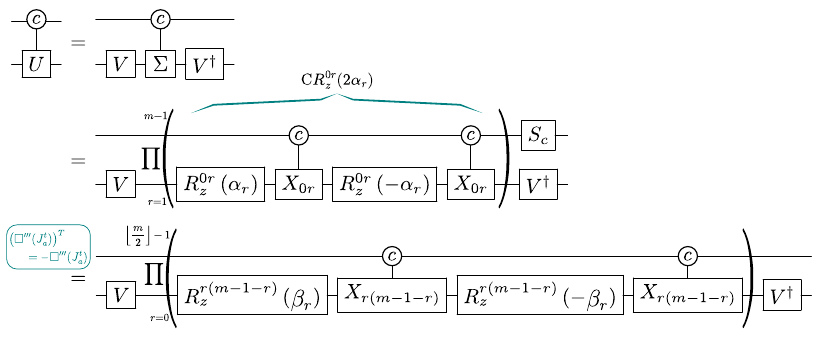}
    \caption{A controlled unitary can be decomposed into two diagonalizing gates and a controlled diagonal operator $C\Sigma$, as presented in Ref.~\cite{Di:2013qvb} (top). The controlled diagonal operator can then be decomposed into a series of controlled $R_z$ operators, each of which can be implemented in two $R_z$ gates and two GCX gates (middle). If $U$ is antisymmetric (as is the case for $\Box'''$), its eigenvalue structure allows the circuit implementation shown in the bottom  diagram, where $R_z$ operators act on pairs of eigenstates with the same eigenvalue up to a sign. In this case, the number of CR$_z$ gates is reduced from $m-1$ to $\lfloor \frac{m}{2} \rfloor$.}
    \label{fig: controlled unitary}
\end{figure}

Compressing a result of Ref.~\cite{Di:2013qvb} within a qudit subspace, the controlled $m$-level unitary at the center of the circuit in Fig.~\ref{fig: multicontrol decomp} can be decomposed into GCX gates by diagonalizing $U = V^\dagger \Sigma V$ as shown in Fig.~\ref{fig: controlled unitary}. 
Let $\Sigma_m$ be the diagonal operator over the $m < d$ dimensional subspace,
\begin{equation}\label{eq: Sigma_m}
    \Sigma_m = \text{diag}\{e^{i\beta_0}, e^{i\beta_1}, ... e^{i\beta_{m-1}}\} = \text{diag}\{e^{i(\varphi - \sum_i \alpha_i)}, e^{i(\varphi + \alpha_1)},e^{i(\varphi + \alpha_2)}, ...e^{i(\varphi + \alpha_{m-1})} \} \ \ \ ,
\end{equation}
where $\varphi = \frac{1}{m} \sum_{i = 0}^{m-1} \beta_i$ and $\alpha_i = \beta_i - \varphi$ for $i\in\left\{1, \cdots, m-1\right\}$.
With $R_z^{jk}(\theta) = \text{exp}\left[ -\frac{i \mathcal{Z}_{jk} \theta}{2} \right]$, where $\mathcal{Z}_{jk}$ is, for the moment, the Hermitian but non-unitary two-level Pauli in the $m$-dimensional subspace, $\Sigma_m$ may be implemented in this subspace as
\begin{equation}\label{eq: Sigma_m in Rz}
    \Sigma_m = e^{i\varphi}R_z^{01}(2\alpha_1)R_z^{02}(2\alpha_2) \ldots R_z^{0(m-1)}(2\alpha_{m-1}) \ \ \ .
\end{equation}
Trivially extending the operators $R_z^{jk}$ to act in the full $d$-dimensional space would accumulate a phase $\varphi$ over the remaining $d-m$ states. For our purposes, those states correspond to gauge variant states, so this constitutes an acceptable GVC. Thus, $\Sigma_m$ can be embedded in the $d$-dimensional space as
\begin{equation}\label{eq: Sigma with phase GVC}
    \Sigma_m \oplus e^{i\varphi}\mathbb{I}_{d-m}= e^{i\varphi}R_z^{01}(2\alpha_1)R_z^{02}(2\alpha_2) \ldots R_z^{0(m-1)}(2\alpha_{m-1}) \ \ \ ,
\end{equation}
where $\mathcal{Z}_{jk}$ has returned to its prior definition in the full $d$-dimensional space.
The controlled diagonal operator can thus be implemented with $m-1$ controlled $R_z$ gates and one controlled phase gate. This equation has been written for the case that the actively mixed levels are localized to the first $m$ levels, but if that is not the case the indices in Eq.~\eqref{eq: Sigma with phase GVC} can simply be reassigned.

Using the GVC described Eq.~\eqref{eq: Sigma with phase GVC}, where the controlled phase $e^{i\varphi}$ is applied to all $d$ levels of the qubit, 
the controlled phase gate can be implemented as a single-qudit phase gate $S_c = \sum_j (1 + \delta_{cj}(e^{i\varphi} -1))\ket{j}\bra{j}$ (as discussed in Ref.~\cite{Di:2013qvb}) over the control register as shown in Fig.~\ref{fig: controlled unitary}. In that case, the phase gate can be parallelized with the diagonalizing $V^\dagger$ gate on the target qudit so that it does not affect the overall circuit depth. Each controlled $R_z$ can be implemented with $2$ single-qudit $R_z$ gates and $2$ GCXs as shown in the middle circuit of Fig.~\ref{fig: controlled unitary}. 

In the $\Box'''$ part of the circuit, the interleaved approach of Fig.~\ref{fig: gmove to gu}, which allows removal of the control on the $J_a^t$ register for each $G(J_a^t)$, naturally connects to the decomposition scheme of Ref.~\cite{Di:2013qvb} where $G(J_a^t)$ is simply the corresponding $V$ for $U = \Box'''(J_a^t)$. 
Finally, the controlled diagonal operators $\Sigma_m$ are the controlled time-evolution operators $e^{i\tau\tilde{\Box}(J_a^t)}$. In this case, an additional reduction in resources can be made using the anti-symmetry of $\Box'''$ (which follows from the fact that it is Hermitian and imaginary). Due to the antisymmetric structure of $\Box'''$, its eigenvalues are $\{\pm \lambda_i\}$ with $\lambda_i \in \mathbb{R}$. So, the diagonalized time evolution operator $\Sigma_m'$ can be expressed as
\begin{equation}\label{eq: Sigma_m'}
    \Sigma_m' = \text{diag}\{e^{i\beta_0}, e^{i\beta_1}, \ldots, e^{i\beta_{m-1}}\} = \text{diag}\{e^{i\beta_0}, e^{i\beta_1}, \ldots, e^{-i\beta_1}, e^{-i\beta_0}\} \ \ \ ,
\end{equation}
where the prime indicates that the eigenvalues are anti-symmetric about 0. In this case, the phase gate vanishes ($\varphi = \frac{1}{m}\sum_{i = 0}^{m-1} \beta_i = 0$) and $\Sigma_m'$ can be implemented in the full qudit space as
\begin{equation}\label{eq: Sigma_m with antisymm reduction}
    \Sigma' = \Sigma_m' \oplus \mathbb{I}_{d-m} = R_z^{0(m-1)}(-2\beta_0)R_z^{1(m-2)}(-2\beta_1) \ldots R_z^{(\lfloor \frac{m}{2}\rfloor - 1) \lceil \frac{m}{2} \rceil}(-2\beta_{\lfloor \frac{m}{2}\rfloor-1}) \ \ \ ,
\end{equation}
as shown in the bottom circuit of Fig.~\ref{fig: controlled unitary}. Here, $R_z$ gates act on pairs of states with antisymmetric eigenvalues. Notice that if $m$ is odd, there is a $0$-eigenvalue term which does not evolve under time evolution, so only $\lfloor \frac{m}{2} \rfloor$ CR$_z$ gates are needed. Furthermore, because there is no phase gate, this expression leaves unphysical states unchanged so that the GVC is identity, rather than a phase.

\begin{table}
    \begin{tabular}{|c|c|c|}
    \hline
     Circuit components & Gates & GCX count \\
    \hline \hline
    
    $\text{C}^{(\ell)}\text{U}_m$ & $2\ell \text{ GCX}, \text{CU}_m$ & $2\ell + 2(m-1)$            \\
    \hline
    
    $\text{CU}_{m}$ &  $2 V_m, \text{C}\Sigma_m$  & $2(m-1)$           \\
    \hline

    $\text{CU}_{m}'$ &  $2 V_m, \text{C}\Sigma_m'$  & $2\lfloor \frac{m}{2} \rfloor$           \\
    \hline
    
    $\text{C}\Sigma_{m}$ &  $(m-1) \text{C}R_z, S_c$   & $2(m-1)$         \\
    \hline

    $\text{C}\Sigma_{m}'$ &  $\lfloor \frac{m}{2}\rfloor \text{C}R_z$   & $2\lfloor \frac{m}{2}\rfloor$         \\
    \hline
    
    $\text{C}R_z$ &  $2 R_z, 2 \text{ GCX}$ & 2            \\
    \hline
    
    $V_m$ &  $V_m$ & 0            \\
    \hline
    
    \end{tabular}
\begin{tabular}{|cc|c|}

\hline
\multicolumn{2}{|c|}{Simulation Strategy}  & GCX \\
\hline
\hline
    non-deformed & Ref.~\cite{Jiang:2025ufg} & $k^4(2(k+1)^4+30)$ \\ 
    
    deformed baseline &  $2(2\xi_{F_{1,2}} + \xi_{F_3} + \xi_G) + \xi_{\tilde{\Box}}$ & $4(8+2k)N_4(k) + 2(6+2k)N_3(k) + 6k N_1(k)$ \\
    
    deformed reduced &  $4\xi_{F_{1,2}} + 2\xi_{F_3} + \xi_{\Box'''}$ & \parbox[t]{8cm}{$\sum\limits_{\ell = 3}^4 \sum\limits_{m = 1}^{\left\lceil \frac{k+1}{2} \right\rceil} 2^{\ell-2} \left( 2\ell + 2 \left( m-1 \right) \right) n_\ell(m,k) +$ \\ $\sum\limits_{m = 2}^{k+1} 2\left\lfloor \frac{m}{2} \right\rfloor  n_1(m,k)  $}\\
\hline
\end{tabular}
\caption{Top: Basic gate decompositions and GCX resources attributed to each. Subscripts $m$ on gates indicate the number of levels the target unitary acts on. As in Eq.~\eqref{eq: Sigma_m with antisymm reduction}, C$\Sigma_m'$ refers to a controlled diagonal operator with antisymmetric eigenvalues. Similarly, CU$_m'$ refers to a controlled antisymmetric operator. Bottom: The GCX cost of a single plaquette Trotter step for each of the decomposition strategies ($d=k+1$). The q-deformed baseline calculations assume $m=d$ for all control sectors in the calculation of $\xi_O$, while the reduced strategy utilizes the distribution of $m$-level unitaries as discussed in Appendix~\ref{Appendix: counting actively mixed} and Eq.~\eqref{eq: weighted depth sums}.}
\label{tab:gcxCalc}
\end{table}
In total, our decomposition scheme has achieved systematic reductions in circuit resources by: 
\begin{enumerate}
    \item accounting for the fact that many of the unitaries constituting the phased F-moves need only mix $m \leq \lceil\frac{k+1}{2}\rceil <d$ qudit levels,
    \item making use of GVC freedom for the phased F-moves to optimize the implementation of controlled unitaries using an embedded version of the diagonalization procedure presented in \cite{Di:2013qvb},
    \item performing time evolution under $\Box'''$ by interleaving control sectors so that the diagonalizing G unitaries $G(J_a^t)$ do not need controlled circuitry,
    \item and using the anti-symmetry of $\Box'''$ to reduce the depth of the corresponding controlled diagonal operator from $m-1$ to $\lfloor \frac{m}{2}\rfloor$ CR$_z$ gates.
\end{enumerate}
The GCX scaling of plaquette time-evolution operator in the non-deformed theory (as reported in \cite{Jiang:2025ufg}), the q-deformed theory without these reductions (as evaluated by our GVCs and direct application of general decomposition techniques to Eq.~\eqref{eq: Trotter depth}), and the q-deformed theory with our reductions are summarized in Table~\ref{tab:gcxCalc}.
The three rows of this table correspond to the three lines of Fig.~\ref{fig: GCX resources}.

For both q-deformed approaches, the number of GCX gates is found to scale as $\mathcal{O}((k+1)^5)$, reflecting the resource complexity's relationship to the 5-qudit active space upon which the largest unit of the circuit—the four controlled F-move—acts. 

Here, we have focused on GCX resource scaling. To address circuit depth, one needs to additionally account for the depth scaling of the $m$-level single-qudit unitaries $V_m$ used to diagonalize a controlled unitary $\text{CU}_m$. Considering the number of degrees of freedom, an arbitrary single-qudit unitary that mixes $m$ qudit levels (such as the diagonalizing $V_m$ gate) can be synthesized with at most $m^2 - 1$ two-level gates. However, in the presence of parallelized two-level rotations or higher-spin native gates~\cite{Clements:2016rkw, Champion:2024wlh} $m$-level single-qudit unitaries can be realized in depth $\mathcal{O}(m)$, leading the overall circuit depth scaling to also be $\mathcal{O}((k+1)^5)$.

\end{document}